\documentclass[11pt,english]{article}
\usepackage[T1]{fontenc}
\usepackage[latin9]{inputenc}
\usepackage[a4paper]{geometry}
\usepackage{amsthm}
\usepackage{amsmath}
\usepackage{amssymb}
\usepackage{graphicx}
\usepackage{setspace}
\usepackage{babel}
\usepackage{hyperref}
\usepackage{tikz}
\newcommand{\slfrac}[2]{\left.#1\middle/#2\right.}

\title{MD Simulation for Head-on Collision of Liquid Nanodroplets Obeying Modified L-J Potential}

\author{Alexander J. Bell\\ University of Leicester} 

\begin{document}

\maketitle
\abstract{
This project models and studies the `head-on' collision of liquid helium nanodroplets within a vacuum, using molecular dynamics simulation techniques. Programs written in MATLAB and C are utilized in tandem to facilitate computer experimentation that achieves this goal. The most expensive computation, that of collision simulation, is handled by a HPC cluster `ALICE' at the University of Leicester, and less expensive peripheral tasks of state initialization and simple analysis are handled by the author's laptop. 

Colliding droplets are modelled as roughly spherical collections of points, cut from a simple cubic lattice, obeying a modified Lennard-Jones potential, with average velocities initialized to ensure a `head-on' collision. These point-sets are then allowed to collide within a cuboid region, designed to take advantage of the observed angular distribution of post-collision fragmentation (favouring a plane orthogonal to `collision axis'), with `translational' boundary conditions that remove vacating points from the simulation region.

To implement the developed theoretical model, an existing C script by D. C. Rapaport, for modelling a homogeneous liquid state, is edited by the author to fit the given, highly heterogeneous scenario. To do analysis on the resulting positions and velocities of points in time, a precise definition of `droplet' was required. An object is defined, from the perspective of metric space, to satisfy this need. An existing cluster analysis algorithm, DBSCAN, is used to apply this definition to the points in simulation (finite subset of $\mathbb{Q}^3$).

The results presented here focus on three distinct properties of post-collision droplets, these being size, speed and temperature; for collision speed varying across ten equidistant averages chosen based on visual examination of collisions. Quantitative evidence of post-collision droplet speed being inversely proportional to droplet radius is presented, droplet temperature distribution post-collision is noted, and qualitative change in collision behaviour across a certain threshold of collision speed is observed.
}
\newpage
\hypertarget{c1}{} \hypertarget{c2}{} \hypertarget{c3}{} \hypertarget{c4}{} \hypertarget{c5}{} \hypertarget{c6}{}
\hypertarget{c7}{} \hypertarget{c8}{} \hypertarget{c9}{} \hypertarget{c10}{} \hypertarget{c11}{} \hypertarget{c12}{}
\hypertarget{c13}{} \hypertarget{c14}{} \hypertarget{c15}{} \hypertarget{c16}{} \hypertarget{c17}{} \hypertarget{c18}{}
\hypertarget{c19}{} \hypertarget{c20}{} \hypertarget{c21}{} \hypertarget{c22}{} \hypertarget{c23}{} \hypertarget{c24}{}
\hypertarget{c25}{} \hypertarget{c26}{} \hypertarget{c27}{} \hypertarget{c28}{} \hypertarget{c29}{} \hypertarget{c30}{}
\hypertarget{c31}{} \hypertarget{c32}{} \hypertarget{c33}{} \hypertarget{c34}{} \hypertarget{c35}{} \hypertarget{c36}{}
\hypertarget{c37}{} \hypertarget{c38}{} \hypertarget{c39}{} \hypertarget{c40}{} \hypertarget{c41}{} \hypertarget{c42}{}
\hypertarget{c43}{} \hypertarget{c44}{} \hypertarget{c45}{} \hypertarget{c46}{} \hypertarget{c47}{} \hypertarget{c48}{}
\hypertarget{c49}{} \hypertarget{c50}{} \hypertarget{c51}{} \hypertarget{c52}{} \hypertarget{c53}{}
\tableofcontents
\newpage

\pagenumbering{arabic}

\section*{\hyperlink{c1}{Introduction}}
\addcontentsline{toc}{section}{Introduction}


The overall goal of this project was to model and study the head-on collision of liquid helium nanodroplets in a vacuum. To demonstrate successful completion of this objective, and give a thorough exposition on the project as a whole, the author has opted to present the work in three distinct chapters; \textbf{Theory}, \textbf{Implementation} and \textbf{Results}, covering non-practical work (ideas and designs for modelling), practical work (essentially programming and model development) and some model results (mainly observational, with basic hypothesis testing) respectively.

It should be noted that the first two sections of implementation focus heavily on demonstrating practical work, altering C code and writing MATLAB code, to show ideas developed in chapter 1 being put into practise. This may be skipped by any reader uninterested in such details, without losing continuity. To any keen programmer, the author would advise looking at `Cell Subdivision Optimization' and `Production of Visuals' within these sections, to see if there's something useful to you there (programming techniques).

There was a great deal of visual material produced during the course of the project; as this guides the author in developing the simulation method, provides a simple means of detecting \href{https://drive.google.com/file/d/0B1syDa_jHh0fT1NhRnFiUjI3MHc/view?usp=sharing}{mistakes in the code${}^{\star}$}\footnote{N.B. $\star$ superscript indicates inconspicuous hyperlink present in the text (in lieu of auto-highlighting)}, and is \href{https://drive.google.com/file/d/0B1syDa_jHh0fRGdtUi1kbUxKNnM/view?usp=sharing}{highly enjoyable to observe${}^{\star}$}. To share this, a link is provided here to some of the more impressive displays (with music references):\\

\begin{footnotesize}
\url{https://drive.google.com/file/d/0B1syDa_jHh0fenFuNUpSeDU5cUk/view?usp=sharing}
\end{footnotesize}
$\;$\\

Additionally, a link to a zipped folder containing all pertinent code, documents and project data is provided here, to serve as both a reference in reading this text, and assist anyone wishing to explore and / or build upon the work presented:\\

\begin{footnotesize}
\url{https://drive.google.com/file/d/0B1syDa_jHh0fcjlXUnJGeWhWajg/view?usp=sharing}
\end{footnotesize}

\section[Theory]{\hyperlink{c2}{Theory}}

The objective of this first chapter is to present theoretical work undertaken on the project; including pertinent background material, modified applications of existing theory, and some ideas developed to meet various requirements / challenges that arose throughout development of the simulation method.\\

\subsection[Relevant Physics]{\hyperlink{c3}{Relevant Physics}}

Simulating colliding nanodroplets (clusters of atoms) is a problem that is most naturally considered from a physical point of view (a physics problem), thus some understanding in this area is necessary to facilitate project work.

\subsubsection[Scientific Method]{\hyperlink{c4}{Scientific Method}}

The general goal of any natural science is to develop `understanding' of some physical phenomena, typically through observation and experimentation. Eventually, a `model' may be proposed, often formulated mathematically, to explain such phenomena by attempting to make predictions based on some available data. As long as no clear contradictions to a model are observed from experiment, it may become accepted as a 'law of nature', such as Newton's laws of motion, relating the concepts of force and acceleration.\\

In modern times, the computer has come to play a central role in this process, as it greatly facilitates the ability to \textbf{simulate} a proposed model, and thus examine the model's results, which must always be distinguished from `real' experimental results. The following diagram, taken from the supervisor's lecture notes \cite[pg.3]{bib:rld1}, inspired by a similar diagram for liquids \cite[pg.5]{bib:alln}, illustrates this process:\\

\begin{center}
\includegraphics[scale=0.65]{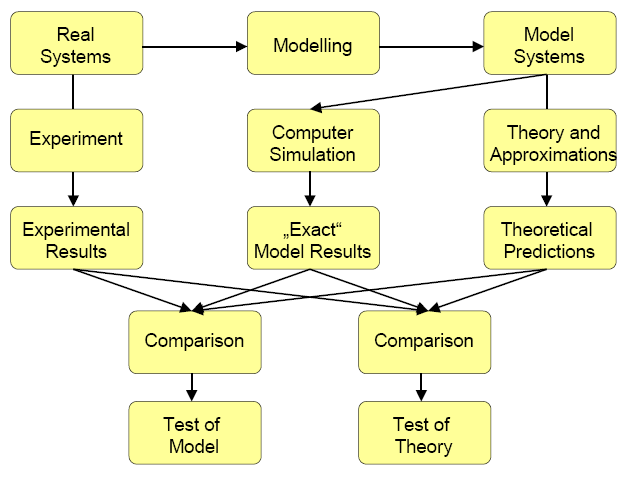}

\textit{fig.1} - how computers interact with the scientific (modelling) process

\end{center}

As it relates to the project, the left branch of fig.1 is essentially ignored, as there is no comparison with real world data. In addition, much of the right branch is ignored as there is no sufficient theory of liquid behaviour for the results to be compared with. Thus it is simply the central path, using classical physics for the model, and pure observation at its end, which this project follows.

\subsubsection[States of Matter and Phase Transitions]{\hyperlink{c5}{States of Matter and Phase Transitions}}

The three common states of matter, solid, liquid and gas, are widely known and experienced concepts to the general public, from relatively early schooling. These describe three sufficiently different states in which collections of atoms / molecules are typically found in nature. The following image, fig.2, will aid understanding:\\

\begin{center}
\includegraphics[scale=0.6]{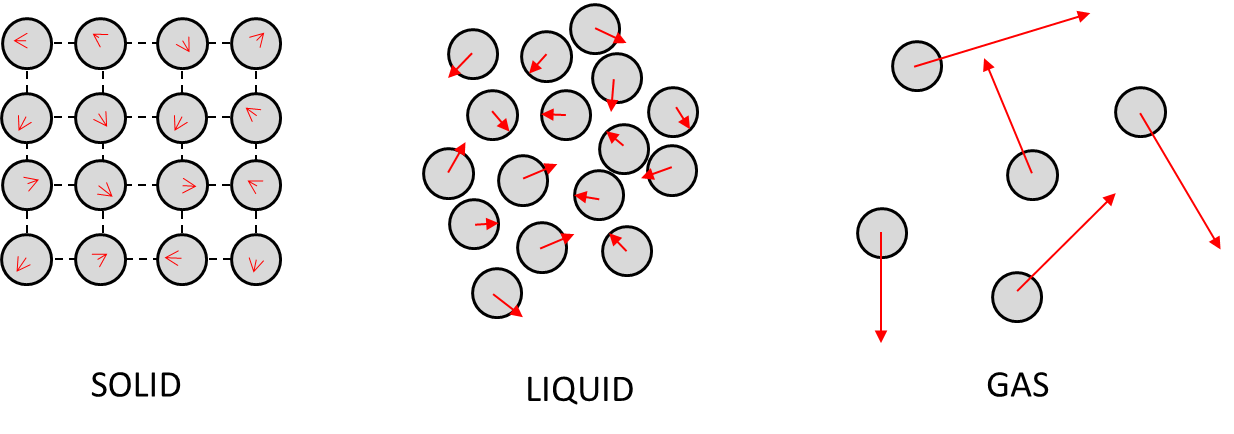}

\textit{fig.2} - behaviour of atoms / molecules in the three common states of matter\\

\end{center}

It is the middle state presented here, liquid, that the simulated nanodroplets must conform to, to be called liquid nanodroplets. This state of matter has proven to be the hardest to describe theoretically, in contrast to the other two, caused by it's lack of inherent structure and dependence on atomic interactions \cite[pg.7]{bib:rapa}.\\

Another term for a "state of matter" is a \textbf{phase} \cite[pg.51]{bib:chem}, and so the term "phase change" refers to a change of state. The main factors that determine the phase of a material, and when it changes, are temperature and pressure. See \hyperlink{A1}{item 1${}^{\star}$}\hypertarget{A1r}{} in the appendix for a diagram of this process as it applies to Helium.\\

In this project, the nanodroplets are propelled through a vacuum, thus the pressure acting an a droplet is practically zero, and it is the temperature of a droplet that will effectively determine its state. Droplet temperature is monitored in model development and compared to model melting point value, which compliments visual examination in justifying the liquid / superfluid state of colliding droplets.

\subsubsection[Temperature Definition]{\hyperlink{c6}{Temperature Definition}}

In a $d$-dimensional system of $N$ atoms, with velocities $v_i \; , \; i = 1,\cdots,N$ , the temperature ($T$) of the system is given by:\\

$$T = \frac{1}{dN} \sum_{i=1}^N |v_i|^2 \;\;\;\;\;\; \cite[\mathsf{pg.15}]{bib:rapa}$$
$\;$

It is desired to measure temperature of a droplet ($D$) travelling within a system (i.e. determine temperature for a subsystem in motion). Suppose $v_1,\cdots,v_{N(D)}$ are the velocities of the $N(D)$ atoms contained within $D$. To take into account relativistic effects, it suffices to subtract the centre of mass velocity ($v(D)$) for the droplet from each $v_i$ in the calculation, to obtain droplet temperature $T(D)$:\\

$$T(D)=\frac{1}{dN(D)} \sum_{i=1}^{N(D)} |v_i-v(D)|^2$$
$\;$

Once droplets have been identified from a system of atoms (to be discussed), this is the equation used to measure their temperature.

\subsubsection[Energy (Kinetic + Potential)]{\hyperlink{c7}{Energy (Kinetic + Potential)}}

Energy is a concept that is central to science \cite[pg.173]{bib:phys}, though is abstract and takes some time (in the author's opinion) to fully grasp. It can be defined as \textit{"the capacity of an object to do work"}, work here being, in one dimension, the product of force and displacement (dot product in higher dimensions).\\

It is possible to consider the total energy an object has, at a given instant, as a sum of two distinct types of energy, kinetic (capacity to do work based on present motion) and potential (capacity to do work based on position in a potential field, such as a gravitational field). Furthermore, it is found that total energy is constant for an isolated system. Consider the 'law of conservation of energy' for a general system \cite[pg.219]{bib:phys}:

$$E_{in}-E_{out} = \Delta E_{sys}$$

In an isolated system, $E_{in}=E_{out}=0$ and so $\Delta E_{sys}=0 \; \Rightarrow \; E_{sys}$ is constant.\\

Consider the following simplified two dimensional system of an idealised pendulum (fig.3), it illustrates how kinetic and potential energy may be exchanged, whilst their sum (by conservation) must remain constant:\\
$\;$

\begin{center}
\includegraphics[scale=0.9]{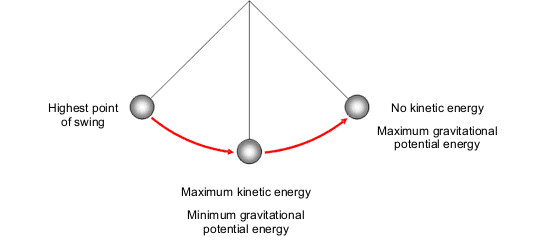}

\textit{fig.3} - pendulum, illustrating an exchange of potential (gravitational) and kinetic energy\\
\begin{tiny}
Source: \url{http://www.bbc.co.uk/schools/gcsebitesize/science/add_aqa_pre_2011/forces/kineticenergyrev2.shtml}
\end{tiny}

\end{center}
$\;$
In this project, energy is used to obtain equations of motion for a given potential function, to be presented shortly, and later to perform a basic check on the validity of simulations (ensure conservation of energy is obeyed, at least approximately). In addition, an energy exchange in collision (Kinetic $\rightarrow$ Potential) will be appreciated.

\subsection[Molecular Dynamics]{\hyperlink{c8}{Molecular Dynamics}}

Molecular dynamics (MD) simulation is, roughly described, a body of material that aims to provide a methodology for computer modelling at the molecular level \cite[preface]{bib:rapa}. It should be noted that, as with all scientific modelling, the result is an \textbf{approximation} to reality, emphasised by the neglect of relativistic effects and quantum principles within its original implementation \cite[pg.5]{bib:rapa}.

\subsubsection[System State]{\hyperlink{c9}{System State}}

A system of $N$ atoms in a particular state (miscrostate) \cite[pg.4]{bib:rld1} at time $t$, may be described by specifying:

\begin{itemize}
\item Simulation region at time $t$, say $R(t)$.\\

In this project, $R$ is constant, set as a cuboid centred at the origin.\\

\item Atom positions at time $t$.\\

e.g. $q(t) = \{\vec{q}_1(t),\vec{q}_2(t),\cdots,\vec{q}_N(t) \} = \{(x_1(t),y_1(t),z_1(t)),\cdots,(x_N(t),y_N(t),z_N(t)) \}$\\

\item Atom linear momenta (mass times velocity) at time $t$.\\

e.g. $p(t) = \{\vec{p}_1(t),\vec{p}_2(t),\cdots,\vec{p}_N(t) \} = \{m_1\dot{\vec{q}}_1,m_2\dot{\vec{q}}_2,\cdots,m_N\dot{\vec{q}}_N \}$\\

\end{itemize}

Thus the system state, at time $t$, may be given by $\{R,q(t),p(t)\}$, where $R$, exemplified in fig.4, is $\{(x,y,z) \in \mathbb{R}^3 \; | \; x \in [-l_x,l_x] \; , \; y \in [-l_y,l_y] \; , \; z \in [-l_z,l_z] \; , \; l_i \in \mathbb{R}_+ \}$.\\

\begin{center}

\includegraphics[scale=0.7]{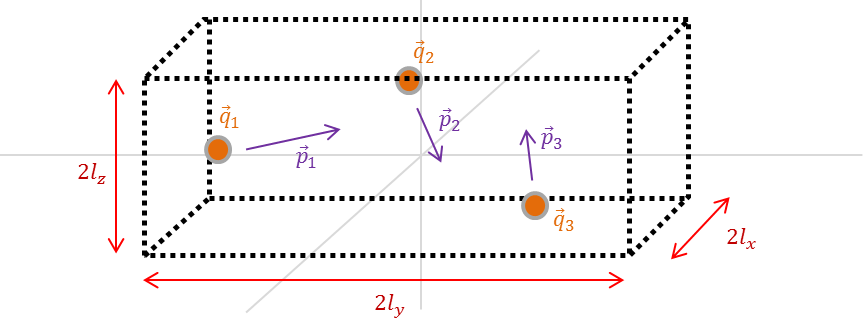}

\textit{fig.4} - simple cuboid simulation region type used in project \\

\end{center}

\subsubsection[Hamiltonian and Potential Functions]{\hyperlink{c10}{Hamiltonian and Potential Functions}}

In classical mechanics, the Hamiltonian ($H$) of a system is the \textbf{total energy} contained within the system, expressed as a function of coordinates $q$ and associated momenta $p$, that is:\\
$$H(q,p) = K(p) + U(q)$$
$\;$

Where $K$ is the total kinetic energy, as a function of the momenta of the atoms, and $U$ is the total potential energy, as a function of the positions of the atoms within the system. Given that an object with mass $m$ and velocity $\vec{v}$ has kinetic energy:
$$\frac{1}{2}m|\vec{v}|^2$$

It is clear to see that the total kinetic energy of a system of $N$ atoms is:

$$K(p) = \frac{m}{2}\sum_{i=1}^N \left\lvert \frac{\vec{p}_i}{m} \right\rvert^2$$
$\;$

Where the mass is constant, thus $\frac{\vec{p}_i}{m} = \frac{m\vec{v}_i}{m} = \vec{v}_i$, and $|...|$ is the Euclidean norm, which in this setting gives the speed of an atom with velocity $\vec{v}_i$.\\

The potential energy of a system may be expressed as sums:
$$U(q) = \sum_1^Nu_1(\vec{q}_i)+\sum_{i=1}^{N-1}\sum_{j=i+1}^Nu_2(\vec{q}_i,\vec{q}_j)+ \cdots$$

In this project, it is only the pair potential ($u_2$ above) that is used, accounting for simple interaction between pairs of atoms. The term with $u_1$ accounts for external forces, which are neglected here (system modelled is isolated), and further sums (over triples, quadruples, and so on...) accounting for more complex interactions are beyond the scope of this project, and computationally expensive to manage \cite[pg.5]{bib:rld1}.\\

The total potential energy may now be expressed $U(q) = \sum_{i=1}^{N-1}\sum_{j=i+1}^N u(r)$, where $r = |(\vec{q}_i-\vec{q}_j)|$, in other words, it is assumed that the potential energy contributed by a pair of atoms depends only on the distance between them. This pair potential can describe the two principal features of inter-atomic force between two atoms, that is resistance to compression (repelling for small $r$) and attraction over a range (for larger $r$) \cite[pg.11]{bib:rapa}.

\subsubsection[Equations of Motion]{\hyperlink{c11}{Equations of Motion}}

To derive equations of motion given a (pair) potential function, the following equations from classical mechanics are used \cite[pg.7]{bib:rld2}, where $k$ indexes degrees of freedom (individual components of the $3N$-dimensional vectors $q$ and $p$):

$$\frac{dq_k}{dt}=\frac{\partial H}{\partial p_k} = \frac{\partial K(p)}{\partial p_k}=\frac{p_k}{m} \;\;\;\;\; , \;\;\;\;\; \frac{dp_k}{dt}=-\frac{\partial H}{\partial q_k} = -\frac{\partial U(q)}{\partial q_k} = -\frac{\partial}{\partial q_k}\sum_{i=1}^{N-1}\sum_{j=i+1}^Nu(r)$$
$\;$

Now given a (pair) potential function $u$, a system of first order ODEs may be obtained, which must be solved numerically using the initial state of the system as the required initial conditions. The numerical method used in this project is a modification of the well known \textbf{leap-frog} method, a simple yet sufficient method of integration \cite[pg.18]{bib:rapa}.\\

\subsubsection[Verlet Method]{\hyperlink{c12}{Verlet Method}}

This project uses the \textit{Verlet method} of numerical integration, which for a system of ODE's...

$$\frac{dq_k(t)}{dt}=p_k(t) \;\;\;\;\; \frac{dp_k(t)}{dt}=f_k(t) \;\;\;\;\; q_k(t=0)=q_{k,0} \;\;\;\; p_k(t=0)=p_{k,0} \;\;\;\; f_k(t=0)=f_k(q_{k,0})$$
$\;$

... prescribes the following algorithm for computing solution at time $t+\Delta t$:

$$\textit{Update velocities by half time-step: } \;\;\; p_k\left(t+\frac{\Delta t}{2}\right)= p_k(t)+\frac{\Delta t}{2} f_k(t)$$
$\;$
$$\textit{Use this intermdiate value for moving: } \;\;\; q_k(t+\Delta t)= q_k(t)+\Delta t p_k\left(t+\frac{\Delta t}{2}\right)$$
$\;$
$$\textit{Compute new force values from new positions: } \;\;\; f_k(t+\Delta t) = f_k(q_k(t+\Delta t))$$
$\;$
$$\textit{Determine new velocities: } \;\;\; p_k(t+\Delta t)= p_k\left(t+\frac{\Delta t}{2}\right) + \frac{\Delta t}{2} f_k(t+\Delta t)$$
$\;$

This description is interpreted from Rapaport \cite[pg.19]{bib:rapa} and the supervisors notes \cite[pg.11]{bib:rld2}. Given an initial state (the initial conditions for the ODE's), this method can be used to compute approximately future states (simulate). The method conserves total momentum, is time-reversible and symplectic \cite[pg.11]{bib:rld2}.

\subsubsection[Dimensionless Units]{\hyperlink{c13}{Dimensionless Units}}

We can define a set of dimensionless units (MD units), by choosing $\sigma$, $m$ and $\epsilon$ to be the units of length, mass and energy, respectively \cite[pg.13]{bib:rapa}. Doing this has a number of benefits, which includes simplifying the equations of motion through absorbing model parameters \cite[pg.13]{bib:rapa} and preventing the use of numbers that lie outside the range of machine precision.\\

e.g. measuring mass in $m$ we have: $p_i=mv_i = v_i$ (in MD units).\\

The effect $\sigma=m=\epsilon = 1$ has in working may be best appreciated in a later section on the particular potential function used in this project, though the reader may perceive a simplification of the first equation of motion (involving $K$) presented previously.

\subsubsection[Boundary Conditions]{\hyperlink{c14}{Boundary Conditions}}

The region of simulation in computation is necessarily finite, and so the problem of dealing with `vacating atoms' must be addressed. \href{https://drive.google.com/file/d/0B1syDa_jHh0fTGxscXhpNU80bEE/view?usp=sharing}{Periodic boundary conditions${}^{\star}$} (PBC) mimic the behaviour of an infinitely large system \cite[pg.16]{bib:rapa}, by reintroducing vacating atoms, component wise (with respect to position coordinates), into the opposite side of the region.\\

PBC were used initially as an existing option in Rapaport's script, however, it was realized that such conditions would interfere with the results of collision simulation (droplets would renter the region and risk `unwanted' additional collisions). Alternative boundary conditions were implemented (see fig. 5), first \href{https://drive.google.com/file/d/0B1syDa_jHh0feDcxOWM0OXNoMXM/view?usp=sharing}{`sticky'${}^{\star}$} (vacating atoms held at the point where they meet the boundary), then \href{https://drive.google.com/file/d/0B1syDa_jHh0fdFV1MWtXaTlkaG8/view?usp=sharing}{`translational'${}^{\star}$} (vacating atoms stored outside region, essentially ignored thereafter), the latter being implemented in the main simulations.\\

\begin{center}

\includegraphics[scale=0.55]{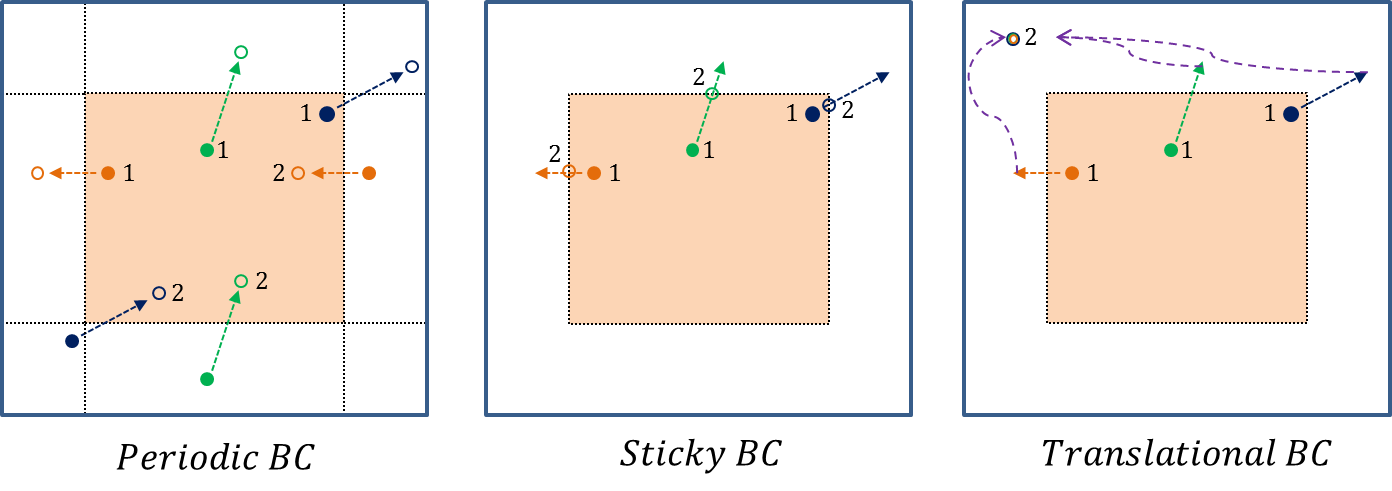}

\textit{fig.5} - demonstration of boundary conditions used during the project \\

\end{center}

\subsection[Lennard-Jones Potential]{\hyperlink{c15}{Lennard-Jones Potential}}

Recall the equation of motion for change in momentum:

$$\frac{dp_k}{dt} = -\frac{\partial}{\partial q_k}\sum_{i=1}^{N-1}\sum_{j=i+1}^Nu(r) \;\;\;\;\; , \;\;\;\;\; r = |\vec{q}_i-\vec{q}_j|$$

An equivalent form from the supervisor's lecture notes \cite[pg.4]{bib:rld1} is as follows:

$$\vec{F}_i = \frac{d\vec{p}_i}{dt} = -\nabla_i\:\sum_{j=1,j \neq i}^{N}u(r) \;\;\;\; , \;\;\;\; \nabla_i = \left( \frac{\partial}{\partial x_i} , \frac{\partial}{\partial y_i} , \frac{\partial}{\partial z_i} \right) \;\;\;\; , \;\;\;\; i \in [1,N]$$

Note the switch back to $i$ for indexing atoms (rather than their components), where $\vec{F}_i$ is the force (equivalently acceleration here, using MD units) on atom $i$, and the new form of the sum may be interpreted as "take the $i^{th}$ atom, and compare with all other $N-1$ atoms", which should be fairly intuitive since $\nabla_iu(|\vec{q}_{a\neq i} - \vec{q}_{b\neq i}|)=0$.\\

The Lennard-Jones potential was an option provided to the author, as it is well known for use in modelling noble gasses \cite[pg.6]{bib:rld1}. Fig. 6 demonstrates the general behaviour of inter atomic interaction for this function. N.B. it is the \textbf{gradient} of the curve that gives force for a given distance of separation, thus it may be observed that atoms repel at close range (small $r$) and attract past a certain point.\\

\begin{center}

\includegraphics[scale=0.5]{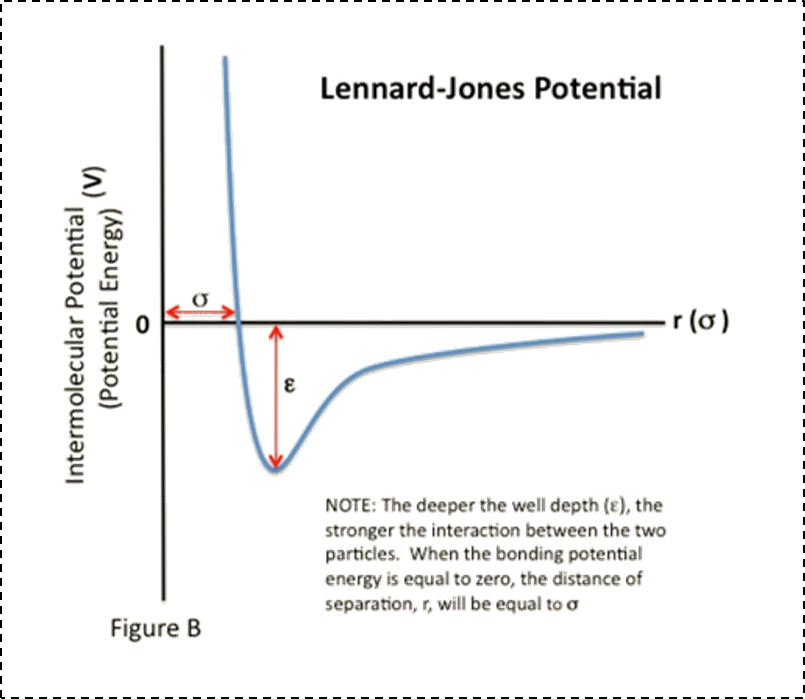}
$\;$\\
$\;$

\textit{fig.6} - graph of Lennard-Jones potential function from Chemwiki \\
\begin{small}
Source: \url{http://chemwiki.ucdavis.edu/@api/deki/files/8914/Figure_B.jpg}
\end{small}

\end{center}

\subsubsection[Function Definition]{\hyperlink{c16}{Function Definition}}

The following is a general definition of the L-J potential function \cite[pg.12]{bib:rapa}:

$$u(r) = 4\epsilon \left[ \left( \frac{\sigma}{r} \right)^{12} - \left( \frac{\sigma}{r} \right)^6 \right] \;\; , \;\; \epsilon \sim \textit{interaction strength} \;\; , \;\; \sigma \sim \textit{defines length scale} $$
$\;$

Use of dimensionless units, described a priori, simplifies this equation to:

$$u(r) = 4(r^{-12}-r^{-6})$$
$\;$

Note that as $r \rightarrow \infty \:$, $\: u^{\prime}(r) \approx 0$, so that the force contribution for atoms sufficiently far apart becomes negligible (very `flat' gradient).

\subsubsection[Modifying the Potential]{\hyperlink{c17}{Modifying the Potential}}

The given potential function can be optimized for computation by modifying the function to be 0 at and past a certain cut-off point, call it $r_c$. To ensure the modified potential function remains 'smooth', an interval $[r_m,r_c]$ is introduced for joining the original potential to the horizontal axis, illustrated in fig.7:

\begin{center}
\includegraphics[scale=0.6]{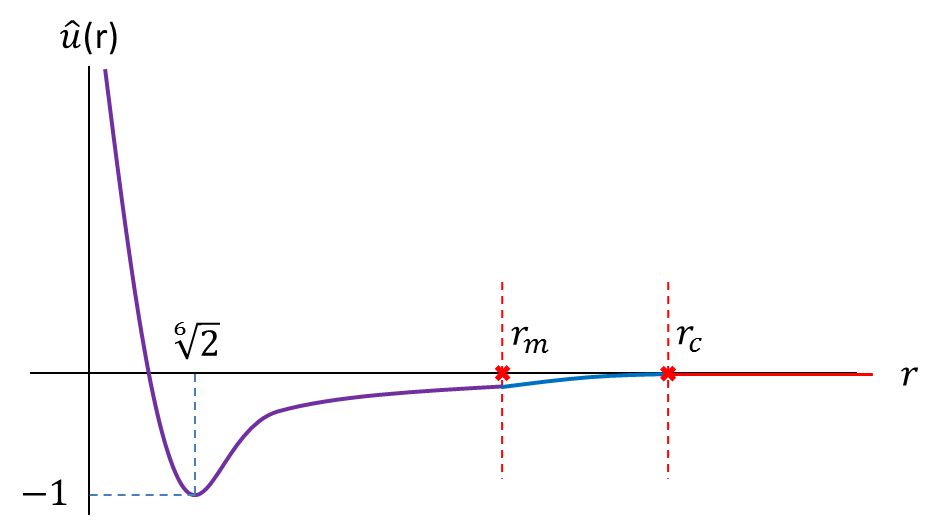}

\textit{fig.7} - modifying the L-J potential function\\

\end{center}

Here $\hat{u}$ is the notation used to denote the modified potential function, distinct from $u$. The question of how one might go about joining $(r_m,u(r_m))$ to $(r_c,0)$ lacks a unique solution. The author's original effort involved using a cubic spline to achieve this goal, which, though technically correct, produces a somewhat complicated and unwieldy expression. A better choice, motivated by the supervisor, is to use a 'switching function', call it $\psi$.

Now the function sought has the following form:\\
\[
    \hat{u}(r)=\left\{
                \begin{array}{ll}
                  u(r) \;\;\; \textit{if } \;\; r \in (0,r_m)\\
                  u(r)\psi(r) \;\;\; \textit{if } \;\; r \in [r_m,r_c]\\
                  0 \;\;\; \textit{if } \;\; r \in (r_c,\infty)
                \end{array}
              \right.
 \]
$\;$

Consider the properties of the function $\psi$. Clearly it is required that:

$$\psi(r_m)=1 \;\;\; , \;\;\; \psi(r_c)=0 \;\;\; , \;\;\; \psi^{\prime}(r_m) = \psi^{\prime}(r_c) = 0$$

Before proceeding further, introduce a change of variables (simplifying later work):

$$x \leftrightarrow \frac{r-r_m}{r_c-r_m} \;\;\;\; , \;\;\;\; r \in [r_m,r_c] \leftrightarrow x \in [0,1]$$
$\;$

So that for a function $ \phi(x(r)) \equiv \psi(r)$, the problem becomes the following:

$$\phi(0)=1 \;\;\; , \;\;\; \phi(1)=0 \;\;\; , \;\;\; \phi^{\prime}(0) = \phi^{\prime}(1) = 0$$
$\;$

To find $\phi$, consider potential candidates for $\phi^{\prime}$. One possibility, the one chosen in this project, is a positive parabola, with roots at $0$ and $1$:

$$\phi^{\prime}(x) = \alpha x(x-1) \;\;\; , \;\;\; \alpha > 0$$

To find $\phi$, simply integrate and use the given boundary conditions, as follows:

$$\phi(x) = \int \phi^{\prime}(x)dx = \alpha\int (x^2 - x) dx = \alpha \left( \frac{x^3}{3} - \frac{x^2}{2} + C \right)$$

$$\phi(0)=1 \;\; \Rightarrow \;\; \alpha C = 1 \;\; \Rightarrow \;\; C=\frac{1}{\alpha} \;\; \therefore \;\; \phi(x) = \alpha\left( \frac{x^3}{3} - \frac{x^2}{2} \right) + 1$$

$$\phi(1)= 0 \;\; \Rightarrow \;\; \alpha \left( \frac{1}{3} - \frac{1}{2} \right) = -1 \;\; \Rightarrow \;\; -\frac{1}{6} \alpha = -1 \;\; \Rightarrow \;\; \alpha = 6$$
$\;$

Thus the following concise form for the modified potential function is realised:
\[
    \hat{u}(r)=\left\{
                \begin{array}{ll}
                  u(r) \;\;\; \textit{if } \;\; r \in (0,r_m)\\
                  u(r)(2x^3-3x^2+1) \;\; , \;\; x = (r-r_m)(r_c-r_m)^{-1} \;\;\; \textit{if } \;\; r \in [r_m,r_c]\\
                  0 \;\;\; \textit{if } \;\; r \in (r_c,\infty)
                \end{array}
              \right.
 \]

\subsubsection[Deriving Forces]{\hyperlink{c18}{Deriving Forces}}

Now that a suitable potential function for computation, $\hat{u}$, has been established; it is required to evaluate $-\nabla_i\hat{u}(r)$ in order to derive equations for force computation \cite[pg.12]{bib:rapa}.

\paragraph{Case 1: $r > r_c$}

Here $-\nabla_i\hat{u}(r) = -\nabla_i 0 = 0$, an expected result.

\paragraph{Case 2: $0 < r < r_m$}

Here $-\nabla_i\hat{u}(r) = -\nabla_iu(r) $. Now to calculate $\nabla_i u(r)$:\\

$$\nabla_i u(r) = \nabla_i \left[ 4(r^{-12} - r^{-6}) \right] = 4 \left[ -12r^{-13} \nabla_i r + 6r^{-7} \nabla_i r \right]$$

$$\nabla_i r = \nabla_i |\vec{q}_i - \vec{q}_j| = \left( \frac{\partial}{\partial x_i} |\vec{q}_i-\vec{q}_j| , \frac{\partial}{\partial y_i} |\vec{q}_i-\vec{q}_j| , \frac{\partial}{\partial z_i} |\vec{q}_i-\vec{q}_j|  \right)$$

$$\frac{\partial}{\partial x_i} |\vec{q}_i-\vec{q}_j| = \frac{\partial}{\partial x_i} \sqrt{(x_i-x_j)^2+ \cdots + (z_i-z_j)^2} = \frac{1}{2} r ^{2^{\left( \slfrac{-1}{2} \right)} } \cdot 2 \cdot (x_i - x_j)$$
$\;$

Thus, using symmetry for other components, $\nabla_i r = r^{-1} (\vec{q}_i - \vec{q}_j)$, and finally:

$$\nabla_i u(r) = 48 (-r^{-14}+ \frac{1}{2} r^{-8})(\vec{q}_i - \vec{q}_j) \;\; \Rightarrow \;\; -\nabla_i u(r) = 48 (r^{-14}- \frac{1}{2} r^{-8})(\vec{q}_i - \vec{q}_j)$$
$\;$

So that for $r<r_m$, the force acting on atom $i$, $\vec{F}_i$, reads:\\

$$\vec{F}_i = 48 \sum_{j = 1, j \neq i}^N \left( |\vec{q_i}-\vec{q_j}|^{-14}- \frac{1}{2} |\vec{q}_i-\vec{q}_j|^{-8} \right)(\vec{q}_i - \vec{q}_j)$$
$\;$

The final case is a little more involved, and will be given abridged.

\paragraph{Case 3: $r_m \leq r \leq r_c$}
$\;$\\

Here $-\nabla_i \hat{u}(r) = -\nabla_i u(r)\psi(r) = -[u(r)\nabla_i\psi(r)+\psi(r) \nabla_i u(r)]$.\\

Now $\nabla_i u(r)$ is known from the previous work, so consider $\nabla_i\psi(r)=\nabla_i\phi(x(r))$:

$$\nabla_i\phi(x) = \nabla_i ( 2x^3 - 3x^2 + 1) = \nabla_i ( 2x^3 - 3x^2) = 3 \cdot 2 x^2 \nabla_i x - 2 \cdot 3x \nabla_i x$$

$$ \nabla_i x = \nabla_i \left( \frac{r-r_m}{r_c-r_m} \right) = \tau \nabla_i r \;\; , \;\; \tau = \frac{1}{r_c-r_m}$$
$\;$

Since $\nabla_i r$ is also known from previous work, an expression for the force $\vec{F}_i$, when the separation distance $r$ is in the 'switching interval' $[r_m,r_c]$, can now be found:

$$\vec{F}_i = \sum_{j=1,j\neq i}^N \left[ 48\left(r^{-14}-\frac{1}{2}r^{-8}\right)(2x^3-3x^2+1)+24\tau x (x-1) r^{-1} (r^{-6}-r^{-12}) \right] (\vec{q}_i - \vec{q}_j)$$
$\;$

Thus a method for calculating force values, given the modified L-J potential function, has been derived. It is the equations presented here that are used in the main program responsible for simulating nanodroplet collisions.

\subsection[`Droplet' Definition and Development]{\hyperlink{c22}{`Droplet' Definition and Development}}

An important question that needs to be addressed in this project is: \textbf{what is a droplet?}\\

To simply say that it is a roughly spherical, perhaps slightly deformed blob of liquid, like an ideal rain droplet for example, is not sufficient for computation / analysis. There appears to be no precise answer readily available to this question, and so the author has opted to form a mathematical definition based initially on the idea of \textbf{connectedness}, rather than shape (see fig.8 for motivation behind this idea).
$\;$

\begin{center}
\includegraphics[scale=0.8]{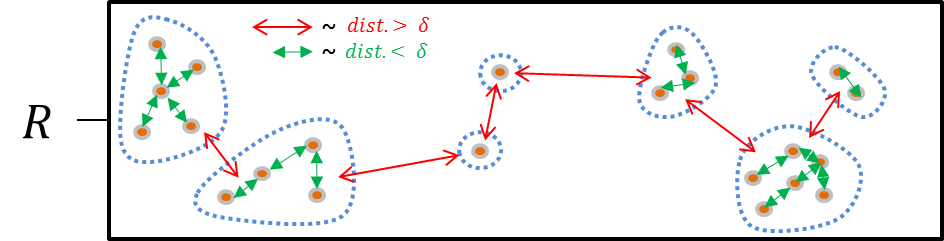}

\textit{fig.8} - motivation for $\delta$-droplet definition\\

\end{center}

\subsubsection[Abstract Setting]{\hyperlink{c23}{Abstract Setting}}

To begin with, a formulation has been sought with the goal of capturing the idea of a set of points being connected (not necessarily continuously), perhaps with some gaps, of maximum value $\delta$. In what follows read $D_{\delta}(x_0 \in X)$ as the $\delta$-droplet in $X$ containing $x_0$.\\

Let $(X,d)$ be a metric space. For any $x_0 \in X$, define $D_{\delta}(x_0) \subseteq X$, $\delta > 0$, as follows:

$$D_{\delta}(x_0) = \bigcup_{n=1}^{\infty} \Omega_\delta^n(\{x_0\}) \;\; , \;\; \Omega_{\delta}(S \subseteq X) = S \cup \{x \in X \setminus S \; | \; d(x,S) \leq \delta \}$$

$$d(x,S) = \textbf{Inf}(\{ d(x,s) | s \in S \}) \;\;\;\; , \;\;\;\; \Omega^n = \underbrace{\Omega \circ \cdots \circ \Omega}_{\textit{n times}}$$
$\;$

This constructive definition bears with it the suggestion of an algorithm for finding a droplet defined in this way. Take a point $x_0$, then include all points within $\delta$ of $x_0$, then include all points within $\delta$ of those points, and so on (guiding vision for the definition).

\subsubsection[Application to Project]{\hyperlink{c24}{Application to Project}}

In this project, the $\delta$-droplet definition can be applied by taking $X = q \subset \mathbb{R}^3$, the atom positions, and $d$ to be the standard Euclidean metric ($d_2$), giving linear distance in space. Thus a basis for identifying droplets from a set of points is established.\\


A natural question to ask now is: \textbf{what should} $\delta$ \textbf{be?} If $\delta$ is too big, there is only one droplet, if it is too small, it simply counts the atoms. Somewhere in between such extremes, a more appealing image of 'droplets' begins to emerge (see fig.9). This project uses $\delta=3$, seen to be sufficient for effective droplet identification (from testing).

\begin{center}
\includegraphics[scale=0.9]{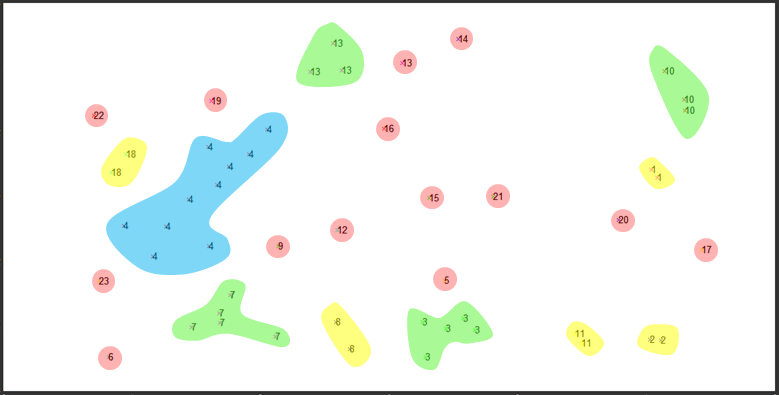}

\textit{fig.9} - identifying $\delta$-droplets in 2 dimensions, annotated output\\

\end{center}

In attempting to put this idea of droplet into practise, identifying droplets from a large (over 20,000) set of points in $\mathbb{R}^3$, it was required to either develop an algorithm for doing this, or find an existing one that facilitates its application.\\

An existing algorithm, DBSCAN \cite{bib:dbscan}, was presented to the author as it applies the given definition in the case of a finite subset of $\mathbb{R}^3$, with the additional flexibility of a second parameter to only include points if they are connected to some minimum number of points. It is a revised version of this algorithm (source to be provide anon) that this project uses.\\

A mathematical description of this more general formulation, using the same preamble as initial definition, only with the constraint that $X$ is finite, and incorporating an additional parameter ($\rho$) to emulate the \textit{minpts} variable from DBSCAN, is as follows:

$$D_{\delta,\rho}(x_0 \in X) = \emptyset \;\;\;\; \textit{if } \;\;\; \#(\Psi_\delta(x_0)) < \rho \;\;\;\; \textit{else}$$
$$D_{\delta,\rho}(x_0) = \bigcup_{n=1}^{\infty} \Theta_{\delta,\rho}^n(\{x_0\}) \;\; , \;\; \Theta_{\delta,\rho}(S \subseteq X) = S \cup \{ x \in X \setminus S \; | \; d(x,S) \leq \delta \: , \: \#(\Psi_\delta(x)) \geq \rho \}$$

$$\Psi_\delta(y_0 \in X) = \{y \in X \setminus \{y_0\} | d(y_0,y) \leq \delta \}  \;\;\;\;\;\;\; \# \sim \textit{caridnality operator}$$
$\;$

Of course this does not yet extend meaningfully to the infinite case, but that is beyond the scope of this project. It should be noted that for a finite set it is now possible to have points not belonging to any droplet, which represents a significant alteration to the previous definition. Thus a view has been adopted that certain clusters of points do not constitute a droplet if they are not large enough (e.g. single atoms and binaries).\\

Again a question arises: \textbf{what should} $\rho$ \textbf{be?}\\

In this project, $\rho = 5$ has been used, as it satisfies excluding the aforementioned undesirable point-sets, and as with $\delta$, has been seen from testing to appear sufficient. Note that $\rho = 0$ is equivalent to the original definition, hence the generalization here.

\subsubsection[State Initialization]{\hyperlink{c25}{State Initialization}}

Before a simulation of droplet collision can be visually appreciated, or even before a collision can take place, there must be a way of initiating a system with droplets properly positioned and velocities set to ensure a collision occurs.\\

To deal with this, the author has designed (with guidance) methods for initiating a system with these properties. Submitting somewhat to the ideal notion of what a droplet is, the initial positions are set in order to be roughly spherical (circular) for a 3D (2D) droplet. The idea, in 3D, for achieving this, is to first create a cubic lattice, of side length $2\gamma$, and then obtain the positions to be plotted as those found within the resulting inscribed sphere (radius $\gamma$). See fig. 10 for a visual overview of the method.\\

\begin{center}
\includegraphics[scale=0.7]{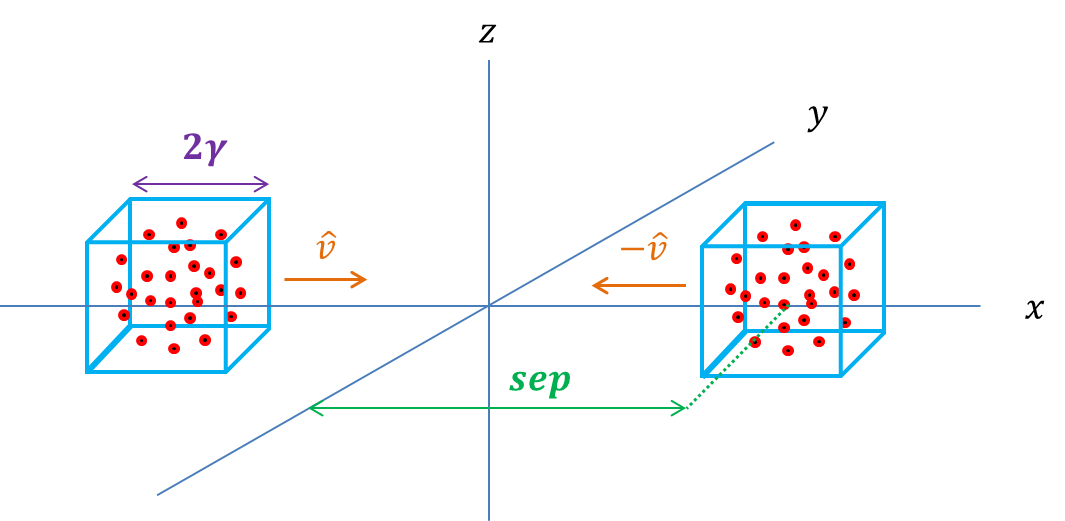}

\textit{fig.10} - droplet initialization in 3 dimensions\\

\end{center}

Now consider a point (atom position) in $\mathbb{R}^3$ within a cubic lattice centred at the origin, test to see if this point lies inside (not including the surface of) the inscribed sphere, and if so, plot two points as translations of the original point, at a distance of the parameter \textit{sep} in the (arbitrarily chosen) positive and negative x directions. This creates two identical `droplets', separated along the x-axis.\\

To encourage a collision, it is sufficient to generate an average initial velocity $\hat{v}$ favouring the positive x-direction, for the droplet in the negative x-direction, as illustrated, and give average initial velocity $-\hat{v}$ to the other droplet. To achieve this, the initial velocity vectors for the atoms belonging to the droplet in the negative x-direction are assigned, using spherical co-ordinates, the values:

$$(\lambda,\theta,\phi) \;\;\; , \;\;\; \lambda \in (\alpha,\beta) \;\;\; , \;\;\; \theta \in ( - \eta, \eta) \;\;\; , \;\;\; \phi \in \left(\frac{\pi}{2} - \eta,\frac{\pi}{2} + \eta \right) \;\;\; , \;\;\; \beta > \alpha \geq 0 \; , \; \eta \geq 0$$
$\;$

Here, the parameters $\alpha$ and $\beta$ govern the initial speed of the atoms, and the parameter $\eta$ allows for the inclusion of some randomness in initial directions, favouring collision. This method is perhaps a little more general than necessary, but the author likes to have options.

The term \textit{`head on' collision} that has been used to describe the nature of the collision being studied in this project can now be precisely defined in the context of this initialization method. Say an initial state is formulated for a head-on collision if $\eta=0$. The Cartesian co-ordinates for the velocities are now provided by the familiar:

$$ x = \lambda\cos (\theta) \sin (\phi) \;\;\; , \;\;\; y = \lambda\sin (\theta) \sin (\phi) \;\;\; , \;\;\; z = \lambda \cos (\phi)$$
$\;$

The co-ordinates for the velocities of atoms in the other droplet are obtained by simply negating the x co-ordinates. Thus a flexible and easily used system for droplet initiation has been provided, with spacing $\sqrt[\leftroot{0}\uproot{2}6]{2}$ used in the initial grid, a choice proven sufficient (though perhaps not ideal) in testing. The means for choosing a droplet size lies in deciding how many atoms should lie on a side of the initial grid, the radius and atom count per droplet are determined from this.

\subsubsection[Spherical Measure]{\hyperlink{c26}{Spherical Measure}}

One final addition to the method of defining `droplet' became necessary towards the end of the project, motivated by the structure of point clusters identified during the collision process. Fig. 11 demonstrates  visual output of two stages in a typical `mid-speed' collision, annotated to show droplet identification given the previously ascribed scheme.

\begin{center}
\includegraphics[scale=0.55]{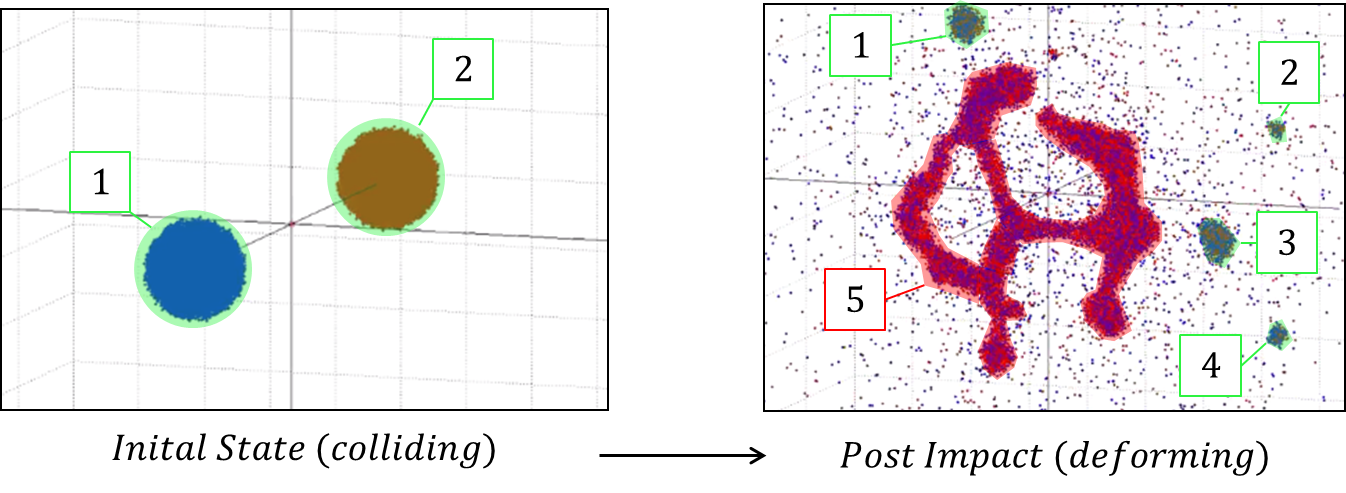}

\textit{fig.11} - droplet deformation post-impact, droplet 5 illustrates `undesirable' identification\\

\end{center}

In wanting to study the speed of droplets \textbf{formed} post collision, the idea that a droplet is `formed' once it has relaxed into a roughly spherical shape emerged, and so a method of measuring the `sphericalness' of a droplet was required. To do this, the author utilized the existing method of state initialization, essentially looking backwards and asking the question, if a droplet contains $N$ points, what is the side length of the simple cubic lattice required to create that droplet? N.B. involves implicit interpolation.

To demonstrate this process, consider a simple cubic lattice of $n^3$ points with spacing $h=\sqrt[\leftroot{0}\uproot{2}6]{2}$. Then the diameter of the inscribed sphere is the side length of this cube, that is $(n-1)h$. The radius of the resulting spherical lattice is thus approximately half this (recall a small thickness of the surface is removed).\\

To approximate the number of points lying within this spherical lattice, consider the ratio of volume of a sphere to its superscribed cube:

$$\frac{Sphere}{Cube} \;\; = \;\; \frac{\frac{4}{3}\pi r^3}{(2r)^3} \;\; = \;\; \frac{4\pi r^3}{4\cdot 3\cdot 2 r^3} \;\; = \;\; \frac{\pi}{6} \;\; \approx \;\; \frac{1}{2}$$
$\;$

Thus for a given number of points in the cubic lattice, we can expect roughly half this number to be contained within the inscribed sphere (accuracy depends on number of points used). However, we implicitly interpolate (assume continuity of parameters, forgetting discrete property of $n$), and say there are $\frac{1}{6}\pi n^3$ points in the sphere.\\

Now reverse engineer the process in order to determine a testing radius $r_t$, to measure how well a droplet $D$ of $m$ points conforms to being a sphere under this algorithm.

$$\#D = m \;\;\; \Rightarrow \;\;\; m = \frac{1}{6}n^3 \pi \;\;\; \Rightarrow \;\;\; n = \sqrt[\leftroot{0}\uproot{2}3]{\frac{6m}{\pi}}$$

$$2r_t = (n-1)h \;\;\; \Rightarrow \;\;\; r_t(m,w) = \frac{w\sqrt[\leftroot{0}\uproot{2}6]{2}}{2} \left( \sqrt[\leftroot{0}\uproot{2}3]{\frac{6m}{\pi}} -1 \right) $$

$\;$

Where $w \in \mathbb{R}_+$ is a weighting parameter, partly governing test stringency. Given a droplet $D$ of $m$ points with centre of mass $C$, loop through all points $d_i \in D$ ($i=1,\cdots,m$), counting the number of times the inequality $|d_i - C| > r_t(m,w)$ is satisfied. Call this number $\mu$. In this project, $w=2$ and $\mu=0$ are used, as this appears sufficient in excluding very `bad' droplets from analysis (determined from experiment).\\


\begin{center}
\includegraphics[scale=0.8]{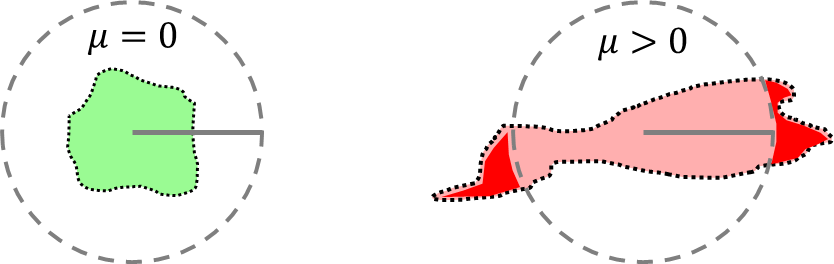}

\textit{fig.12} - spherical measure by counting points lying outside of determined sphere\\

\end{center}


\section[Implementation]{\hyperlink{c27}{Implementation}}

The main purpose of this chapter is to demonstrate putting the material presented previously into practise, which is essentially the bulk of the practical work carried out in this project (programming). The source code for collision simulation is discussed, in particular the editing of this code to meet project requirements.\\

Model testing and development is demonstrated, observing energy conservation / exchange, monitoring droplet temperature, and testing maximum force values for varying time-step sizes. The main simulations, providing raw data for analysis, are described, and the programs written for the results presented in the final chapter are discussed.

\subsection[Simulation Code (C)]{\hyperlink{c28}{Simulation Code (C)}}

The author was advised, early on in the project, to make use of existing material for handling the complex task of efficiently simulating an MD experiment. The source code was thus taken from programs made available by Prof. D. C. Rapaport in support of his book \cite{bib:rapa}. The link for accessing these files is given here:\\

\url{http://www.ph.biu.ac.il/~rapaport/mdbook/getmdsw.php}\\

The code is written in the C programming language (low level), which the author was required to learn, to a basic level, in order to make changes (using `Code::Blocks') that would implement the desired scenario of colliding nanodroplets, supplanting the existing intent which was of a \href{https://drive.google.com/file/d/0B1syDa_jHh0fcnVYa1QzVC1yd3c/view?usp=sharing}{significantly different nature${}^{\star}$}; this being, for the scripts examined, studying thermodynamic properties of a homogeneous distribution of atoms approximating liquid behaviour by a (repulsive only) Lennard-Jones potential.

Though a number of Rapaport's scripts were worked with during the project, the main script used as a basis for the model, to be developed and tested, is \textbf{pr\_03\_2.c} (*). This script is described from page 54 of the book \cite{bib:rapa}, and implements a \textit{cell-assisted neighbour list} method, a technique of force calculation (typically most expensive task) which provides a significant performance boost over other scripts (including attempts made by the author) that rely on the simple `all-pairs' method for calculating force.\\

\subsubsection[Reading Initial State]{\hyperlink{c29}{Reading Initial State}}

In (*), the initial state is generated by functions \textit{InitCoords}, \textit{InitVels} and \textit{InitAccels}, setting atom position, velocity and acceleration respectively. It creates a 3D grid of points determined by the desired grid size and density (input param.), with randomized velocities scaled according to desired temperature (input param.), and accelerations set to zero.\\

The state initialization algorithm, described in chapter 1, explicitly determines positions and velocities, and its implementation (to be discussed) produces text files that must be read in (*), to provide the desired heterogeneous initial state set for a collision. Fig. 13 demonstrates editing \textit{InitCoords}; \textit{InitVels} undergoes an analogous change.\\

\begin{center}
\includegraphics[scale=0.6]{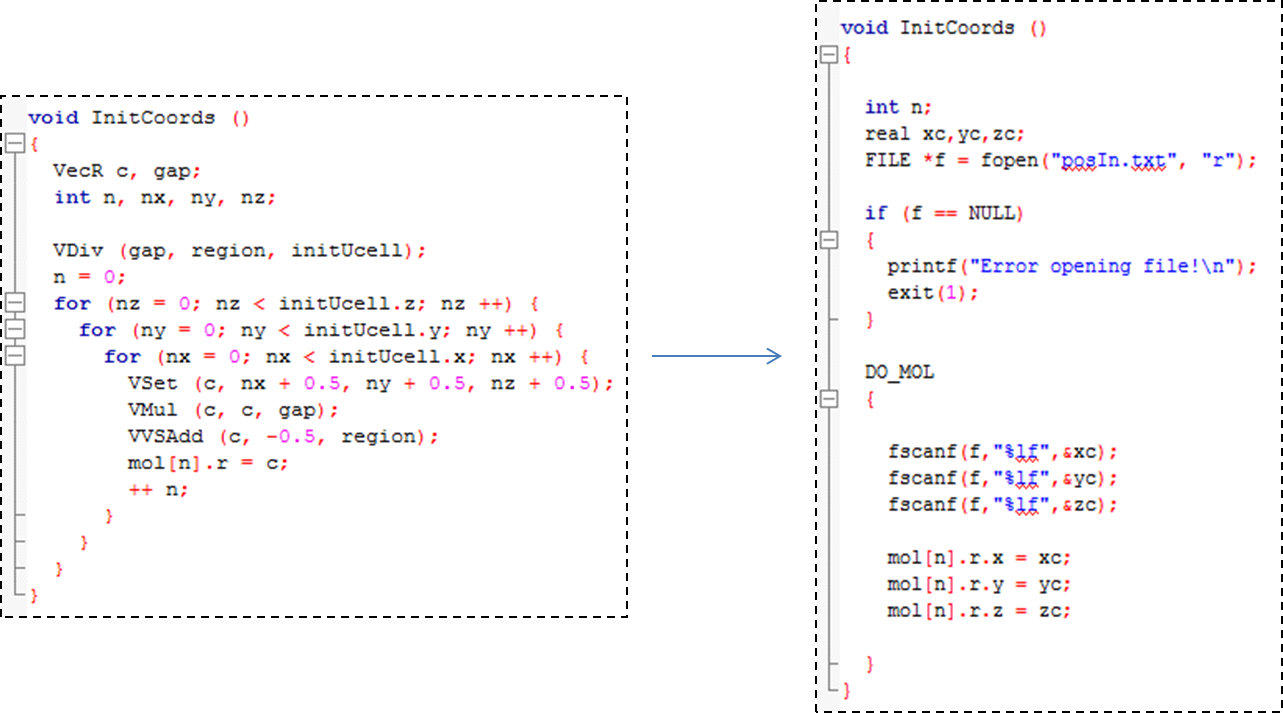}

\textit{fig.13} - adapting state initialization code to read from generated .txt file\\

\end{center}

The acceleration initialization function is left unchanged (accelerations all set to zero). The code for reading from a text file is based on existing material circulating the internet, see: \url{ http://www.programiz.com/c-programming/examples/read-file}.

\subsubsection[Force Modification]{\hyperlink{c30}{Force Modification}}

The force calculation in (*) is handled by the function \textit{ComputeForces}, and as previously described, implements a greatly simplified version of the Lennard-Jones potential which ignores attractive force and has a cut-off point at the potentials minimum ($r=\sqrt[\leftroot{0}\uproot{2}6]{2}$), achieved simply by translating the graph one unit in the positive vertical direction.\\

The code alteration is too extensive to give a `before-and-after' shot here, so the author has created the digram in fig. 14 to exemplify the essential features of the transformation, of the code within the nested loop in \textit{ComputeForces} responsible for force evaluation.\\

\begin{center}
\includegraphics[scale=0.55]{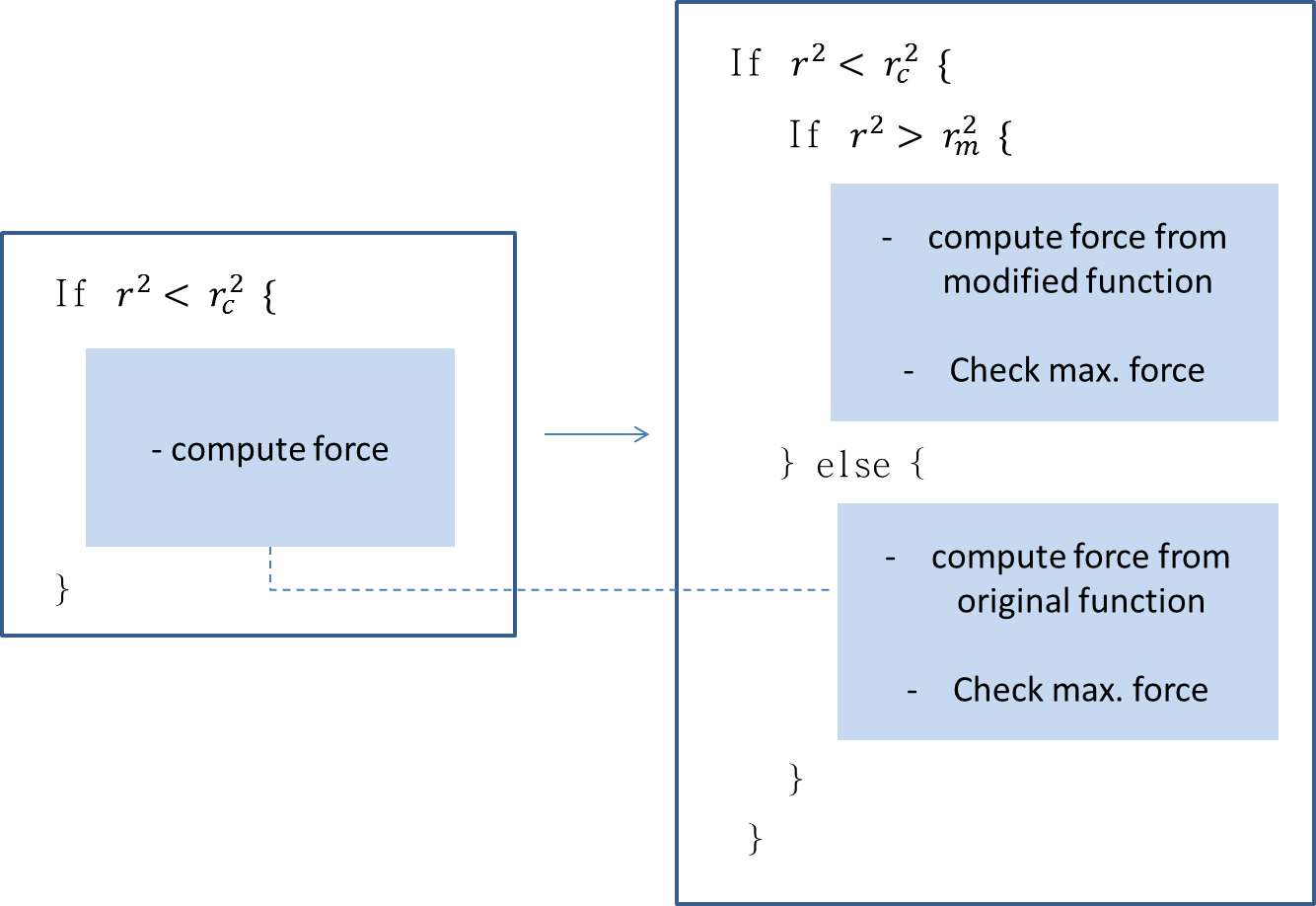}

$\;$\\

\textit{fig.14} - modifying force calculation to implement modified potential and track max force\\

\end{center}

The addition of the `if...else...' statement serves to identify whether or not a valid distance (less than the cut-off point $r_c$) falls within the switching interval ($[r_m,r_c]$) previously derived, in order to assign the right expressions for force calculation and potential energy sum. Note the use of squares in distance comparison, a simple technique to optimize computation speed which the author has attempted to imitate in this alteration.\\

In addition to implementing the switching interval, the new code is modified to observe the maximum force value occurring each time-step, to be presented later in this chapter.

\subsubsection[Altering Boundary Conditions]{\hyperlink{c31}{Altering Boundary Conditions}}

The task of implementing alternative boundary conditions (sticky / translational) proved to be perhaps the hardest alteration to make, requiring some fundamental changes to the existing code. The initial boundary conditions in (*), PBC, are implemented by the function \textit{ApplyBoundaryCond}, which in turn calls to a number of \textit{VWrap} definitions in a local header (.h) file, enforcing the PBC.\\

The modification process will be given in stages. Initially it was suggested that a `flag' variable (integer acting as boolean, initialized to zero) be given to all atoms (altering their `struct' definition), stating whether or not they have left the simulation region, see fig. 15.\\

\begin{center}
\includegraphics[scale=0.5]{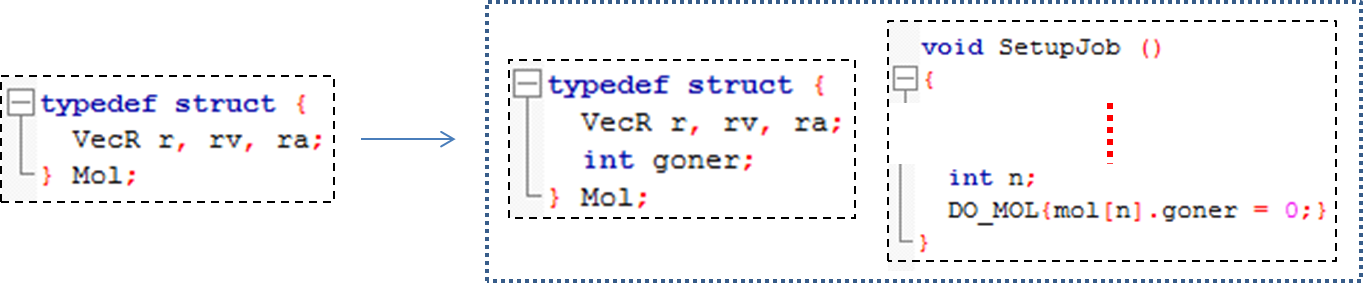}

$\;$\\

\textit{fig.15} - implementing `goner' flag variable and setting to zero in \textit{SetupJob} function\\

\end{center}

Then the \textit{ApplyBoundaryCond} function is altered to check all atoms that aren't `goners' (goner$=0$), determining whether they have left the region, and if so, flagging them (set goner$=1$). In addition, for translational boundary conditions, atom position is respecified to a pre-set point outside of the simulation region, see fig. 16. \\

\begin{center}
\includegraphics[scale=0.5]{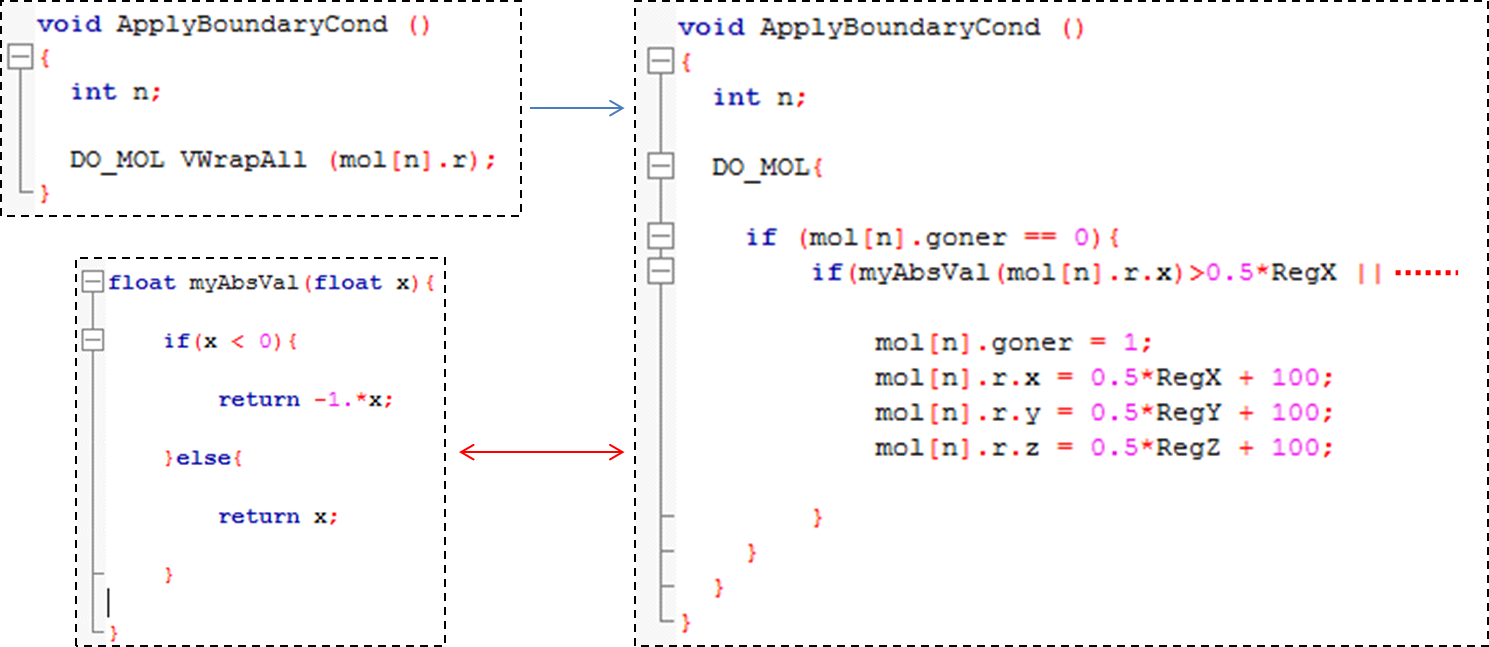}

$\;$\\

\textit{fig.16} - changing boundary condition application from periodic to translational\\

\end{center}

In addition to the demonstrated modifications, a number of functions required editing to become sensitive to the `goner' variable, with an aim to stop a vacated atom (goner$=1$) from interacting with the system. To name two important edits, the \textit{ComputeForces} function (outside of force evaluation) to stop force calculation for `goners', and the \textit{LeapfrogStep} function (stages 1,2 and 4 of Verlet method) to prevent moving these atoms.

\subsubsection[Cell Subdivision Optimization]{\hyperlink{c32}{Cell Subdivision Optimization}}

The \textit{cell subdivision} method for optimizing computation, in certain situations, is a simple but very powerful technique that the author gained a lot of experience in when trying to implement his own droplet identification scheme.\\

The technique can be easily demonstrated for the following 2D problem: Given a finite set of points $S \subset \mathbb{R}^2$ , calculate for all pairs $(s_i,s_j) \in S \times S, i\neq j$, the function $f: S \times S \rightarrow \mathbb{R}$ , $f = f(d_2(s_i,s_j))$, where for some critical value $\epsilon > 0$, $f(r > \epsilon)$ is unimportant. I.e. it is only necessary to apply $f$ to points within $\epsilon$ of each other by the $d_2$ metric.\\

To apply the cell subdivision method to this problem, let $C$ be the smallest set of the form $[l,l+\gamma] \times [l,l+\gamma]$ , $l \in \mathbb{R}$ , $\gamma \in \mathbb{R}_+$ , containing $S$ and minimizing $l$ (a minimal and unique square region containing all points of $S$). Now calculate $\omega(l)=\lceil \frac{l}{\epsilon} \rceil$ and consider a grid of $\omega^2$ cells centred on $C$ with cell length $\epsilon$. Then each $s \in S$ lies within a cell if we adopt some convention for the cell boundaries (e.g. include the left and top edges, leaving others open with obvious exceptions of cells on the right and bottom boundary of $C$).\\

Now for some given $s \in S$ contained in a cell $C_s$, it is only necessary to check points $s_i \in S , s_i \neq s$ belonging to $C_s$ and its adjacent cells, see fig. 17. In any space large and numerous enough, this reduction in checking provides a major saving in computational cost. The reader may perceive that a proper mathematical formulation of this technique would be far more general and detailed than what has been given here.\\

\begin{center}
\includegraphics[scale=0.55]{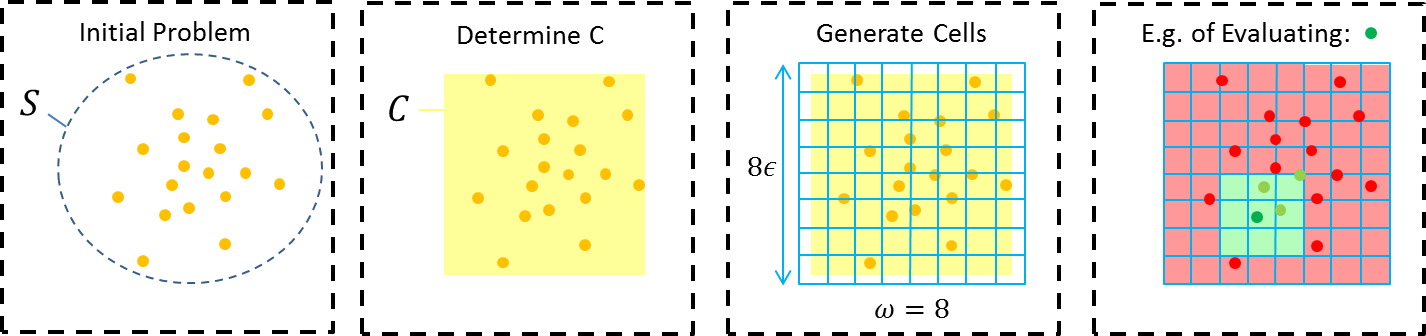}

\textit{fig.17} - demonstrating construction and application of cell subdivision method\\

\end{center}

Suppose given the critical distance $\epsilon$, a distinction $\epsilon_{cell}$ is made for constructing the grid of cells. Obviously $\epsilon_{cell}=\epsilon$ presents no change to the method, and $\epsilon_{cell} < \epsilon$ makes the approach invalid (will risk missing points that are within $\epsilon$ distance). It is clear, however, that for $\epsilon_{cell} > \epsilon$, the method remains valid (will not miss necessary checks). The reason why one might consider such a generalization will be motivated by considering the method applied to the heterogeneous 3D system simulated in this project (see fig. 18).

\begin{center}
\includegraphics[scale=0.55]{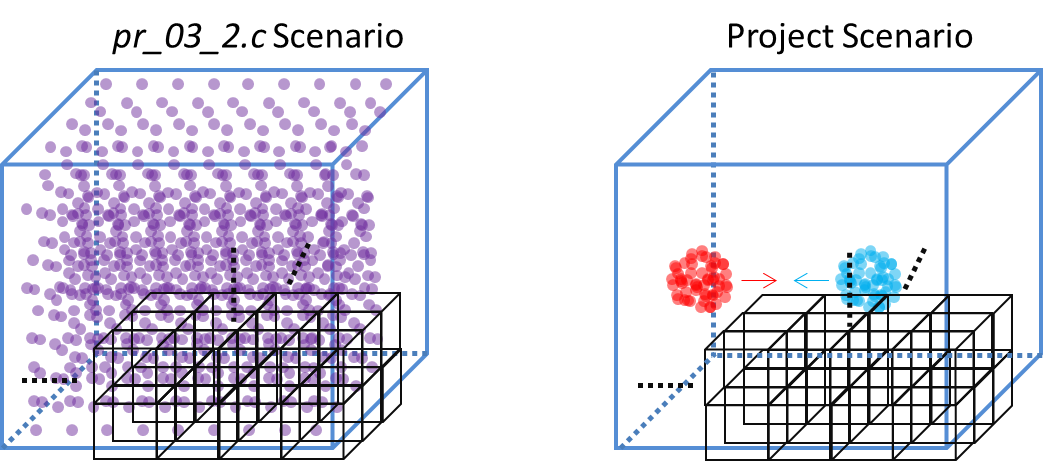}

$\;$\\

\textit{fig.18} - visualizing 3D cell grid for \textit{pr\_03\_2.c} and project simulation scenarios\\

\end{center}

It was desired to make the region size sufficiently large so as to contain all significant collision debris for any reasonable length of time. Unfortunately, the cell subdivision method described becomes very expensive for large regions, as the program must check every cell on iteration (easily reached billions of cells whilst failing to contain collision aftermath). This was using the standard $\epsilon = r_c$ in the grid construction.\\

In developing an algorithm to identify droplets from a large space, the author ran into this same problem, and tried using $\epsilon_{cell} > \epsilon$, witnessing a significant saving in computation time (see fig. 19). N.B. applies well here due to abundance of low density / empty space. 

\begin{center}
\includegraphics[scale=0.45]{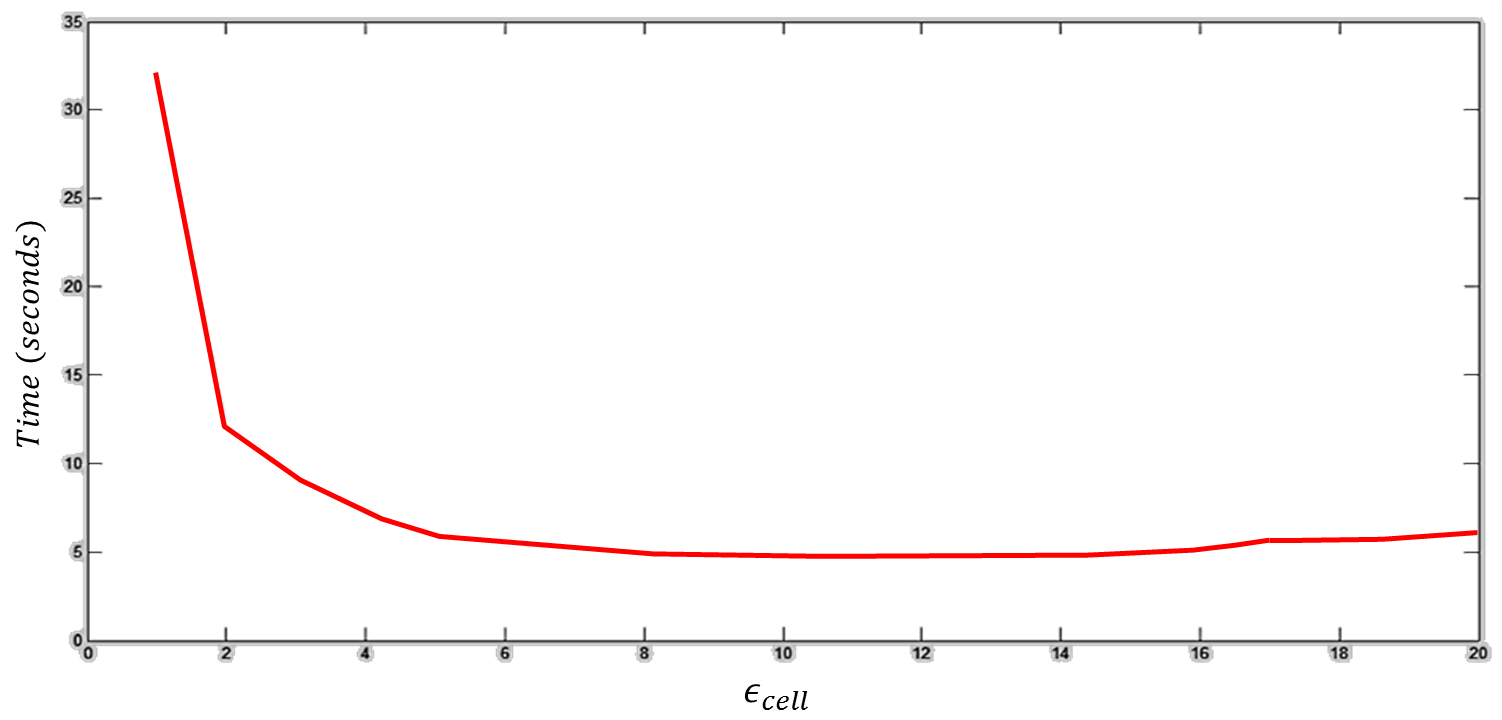}

\textit{fig.19} - computation times of droplet ID for a fixed point-set, increasing $\epsilon_{cell}$\\

\end{center}

This method of increasing the parameter used in cell construction was then applied to the simulation code under development, by introducing the input variable $rCell$ to the code, with the constraint (presently unenforced) that $rCell \geq rCut$ ($r_c$ variable), see fig. 20. The main simulations used $(rCut,rCell)=(3,8)$, chosen simply for sufficiency, and providing a significant reduction in simulation time.

\begin{center}
\includegraphics[scale=0.5]{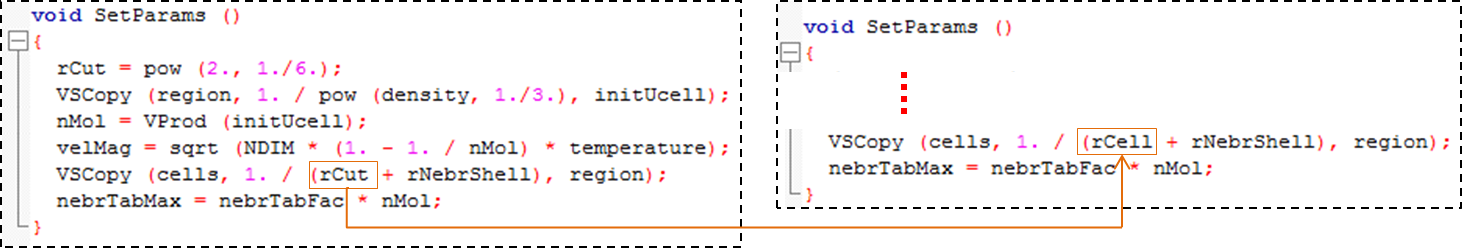}

\textit{fig.20} - implementing cell subdivision optimization in pr\_03\_2.c\\

\end{center}

\subsubsection[Output (writing) Functions]{\hyperlink{c33}{Output (writing) Functions}}

To perform analysis on simulation results, beyond the readout of system properties provided in (*), it was required to have the code write various quantities to text files (storing output for later analysis) at periodic points in execution (every so many time-steps). To accomplish this task, the author wrote a number of additional functions (one for each .txt file) and introduced parameters \textit{photoAvg} and \textit{measAvg} to control periodicity of output for visual and quantitative analysis respectively, see fig. 21 for demonstration.

\begin{center}
\includegraphics[scale=0.52]{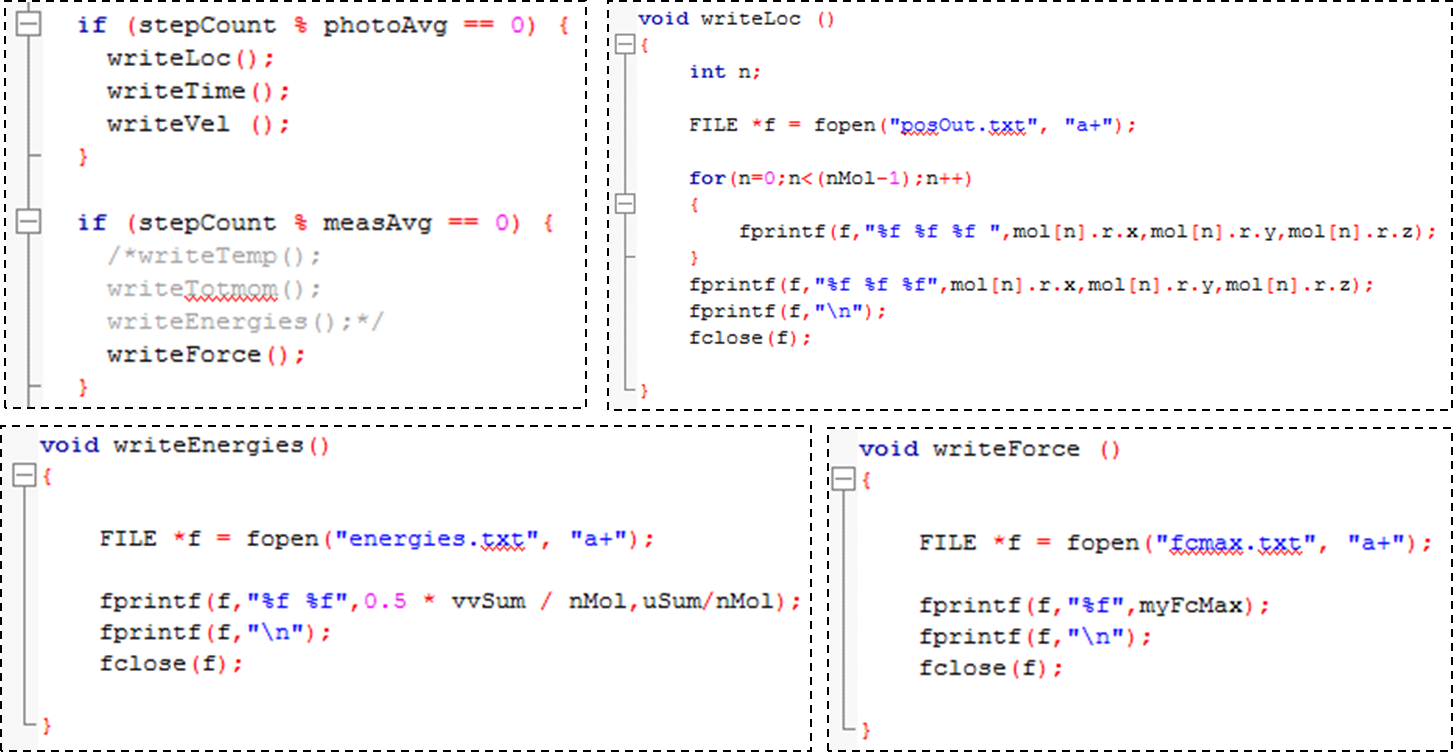}

\textit{fig.21} - functions added to pr\_03\_2.c for outputting values of interest to .txt files\\

\end{center}

The text files produced by these functions form the basis for post-process analysis, e.g., outputting position and time (later realized to be derivable from row count of any output array) values produces the raw data necessary to plot the system.

\subsection[Peripheral Tasks (MATLAB)]{\hyperlink{c34}{Peripheral Tasks (MATLAB)}}

It was decided early on in the project to assign the tasks of state initialization and analysis of simulation output to scripts written in MATLAB, a higher level language; as the computation cost in these areas is far lighter than in the main simulation, and the author's programming experience with MATLAB is significantly greater.

\subsubsection[Generating Initial State]{\hyperlink{c35}{Generating Initial State}}

To implement the state initialization algorithm described in the first chapter, the script \textit{genDrops3D.m} was written, in addition to auxiliary functions \textit{writeDrops.m} and \textit{sphToRect.m}, for writing the generated state to .txt files and converting spherical coordinates to Cartesian equivalents respectively. The main script is provided as \hyperlink{A3}{item 3${}^{\star}$}\hypertarget{A3r}{} in the appendix.\\

The produced .txt files are named \textit{posIn.txt} and \textit{velIn.txt}, storing atom positions and velocities respectively. These files are transferred from the local folder, via SCP protocol, to a folder containing a compiled C script for execution (see fig.22). SCP is also used for transferring output files, from ALICE, to the author's laptop for post-processing.

\begin{center}
\includegraphics[scale=0.55]{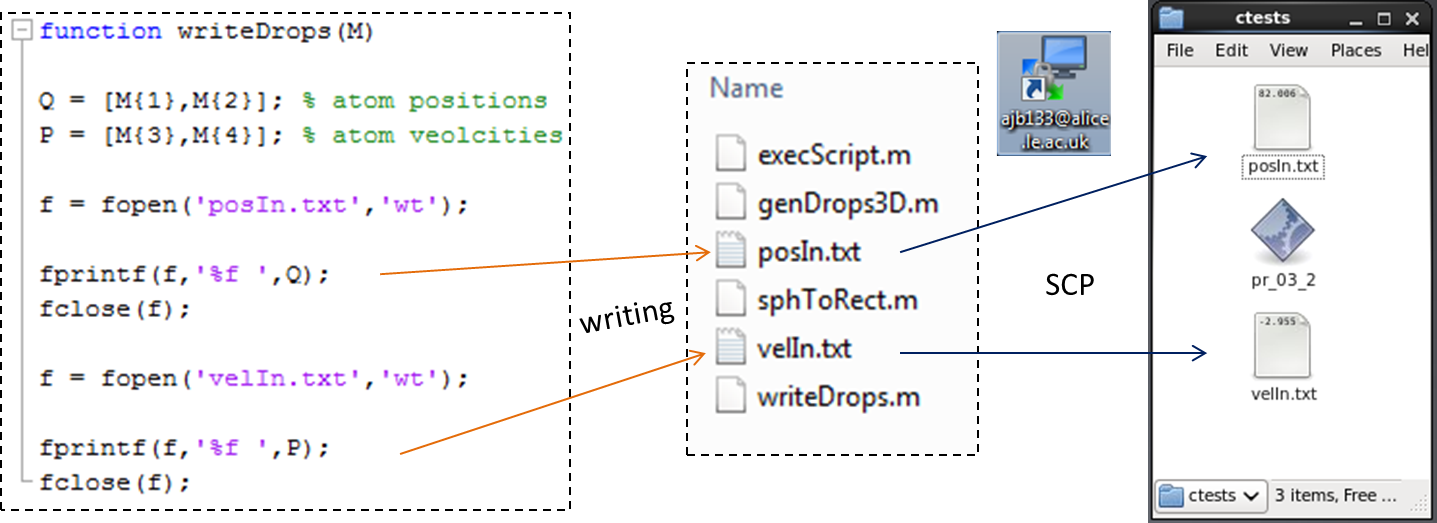}

$\;$\\

\textit{fig.22} - writing initial state from MATLAB and transferring to ALICE for simulation\\

\end{center}


\subsubsection[Production of Visuals]{\hyperlink{c36}{Production of Visuals}}

The ability to \href{https://drive.google.com/file/d/0B1syDa_jHh0fOXFQSFgtVVY1TWM/view?usp=sharing}{\textbf{see} droplet collision${}^{\star}$} has been a strong interest for the author throughout the project. To accomplish this task, a number of MATLAB scripts have been developed, taking atom positions in time from \textit{posOut.txt} (written during simulation), occasionally with other measurements for monitoring with visual state, and processed into a `movie'; predominantly using built-in functions \textit{plot3}, \textit{getframe} and \textit{movie2avi}.

A complete description of the filming process is too extensive to present here; instead two significant features will be demonstrated and discussed briefly, that is, some `dynamic' filming (encoded in the main script \textit{pFilm3D.m}) and simulation region plotting (part of auxiliary function \textit{plotRapPT.m}). See fig. 23 for reference.
\begin{center}
\includegraphics[scale=0.52]{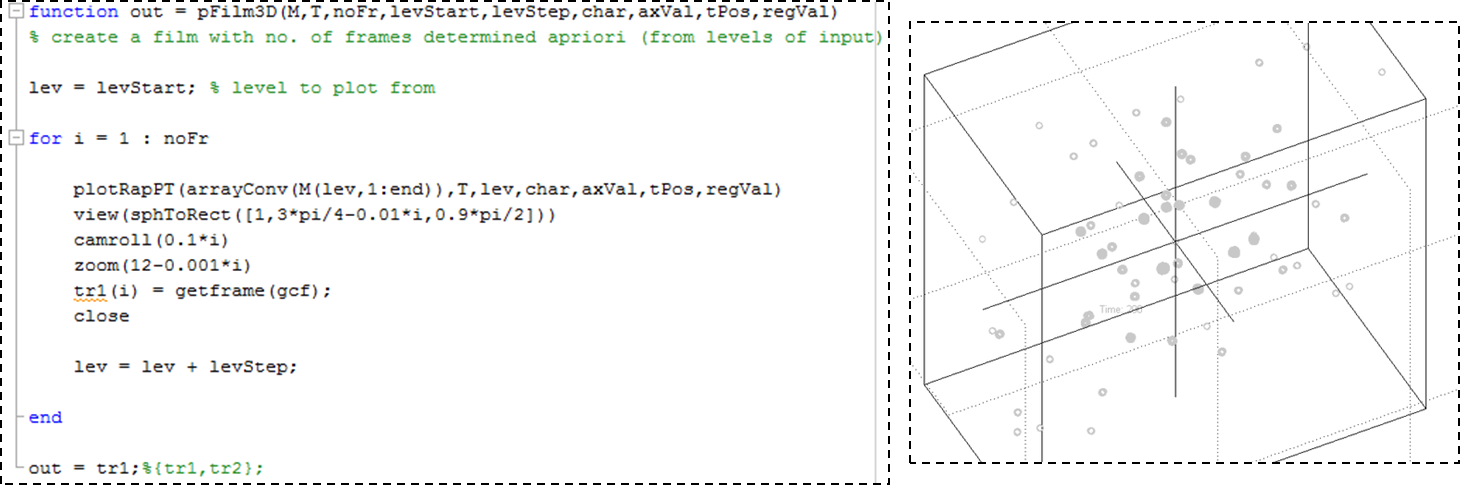}

\textit{fig.23} - dynamic filming script (left) and image of simulation region plotting (right)\\

\end{center}

To really appreciate (visually) the \href{https://drive.google.com/file/d/0B1syDa_jHh0fQzIxY2ROZFB5M0k/view?usp=sharing}{complex and evolving process${}^{\star}$} of droplet collision taking place in simulation, it was vital to have as much control over the filming environment as possible, beyond the simple static view; one wishes to see these phenomena unfolding at \href{https://drive.google.com/file/d/0B1syDa_jHh0fNWhsbmFTSUxmeXM/view?usp=sharing}{multiple levels (zoom) and varying angles (rotation)${}^{\star}$}. \\

Combining the built-in functions \textit{view}, \textit{zoom} and \textit{camroll}, with a dynamic variable tracking frame number (`$i$' in the above script) provides this functionality. Further splitting the process into various stages of different motions can be achieved with careful application of conditional statements. The reader may also recognise use of spherical coordinates, which simplify control of viewing angle.\\

The flexibility of stating precisely the axis range for plotting is provided in \textit{pFilm3D.m} by the \textit{axVal} input parameter ($1\times 6$ array passed to \textit{axis} function in \textit{plotRapPT.m}). To visualize the actual simulation region (independent of plotted region), an additional input parameter \textit{regVal} is introduced. This passes to the plotting function the required values (region lengths) for outlining the simulation region with lines connecting its vertices (total of 12 \textit{plot3} commands), in addition to standard axis plotting.\\

Films generated as frame arrays in MATLAB were exported to .avi using the \textit{movie2avi} command, then post-processing was carried out using freeware \textit{VirtualDub}, available from: \url{http://www.virtualdub.org/}. Merging of videos was performed using \href{http://forum.videohelp.com/threads/310303-Combining-2-Videos-into-1-file-with-two-windows}{.avs script${}^{\star}$}.

\subsubsection[Droplet Identification]{\hyperlink{c37}{Droplet Identification}}

Given atom positions (\textit{posOut.txt}) and velocities (\textit{velOut.txt}) from simulation, it was required to partition such data into `droplets', implementing the theory of what a droplet is, presented in chapter one. To do this, the author made a \href{https://drive.google.com/file/d/0B1syDa_jHh0fRGh6OGZaamFYQXM/view?usp=sharing}{personal attempt${}^{\star}$} to write code identifying droplets, whilst learning to use the DBSCAN algorithm \cite{bib:dbscan} for comparison. The author was unable to outperform the DBSCAN approach (by a factor of minutes), and so it is this algorithm (link below), that forms the basis of droplet identification in this project.

\begin{footnotesize}
\begin{center}
\url{http://www.mathworks.com/matlabcentral/fileexchange/48120-revised-dbscan-clustering}\\
\end{center}
\end{footnotesize}

The script \textit{getDropsPV.m} was written to harness DBSCAN, sorting a given set of positions and velocities into droplets; of course it is only the positions of atoms that are required for droplet ID, the velocities are included here as their droplet association is required for later analysis. Fig. 24 presents this script, and visual output, identifying droplets individually, from \textit{plotDrops.m}, inspired by popular approach used in entomology.

\begin{center}
\includegraphics[scale=0.55]{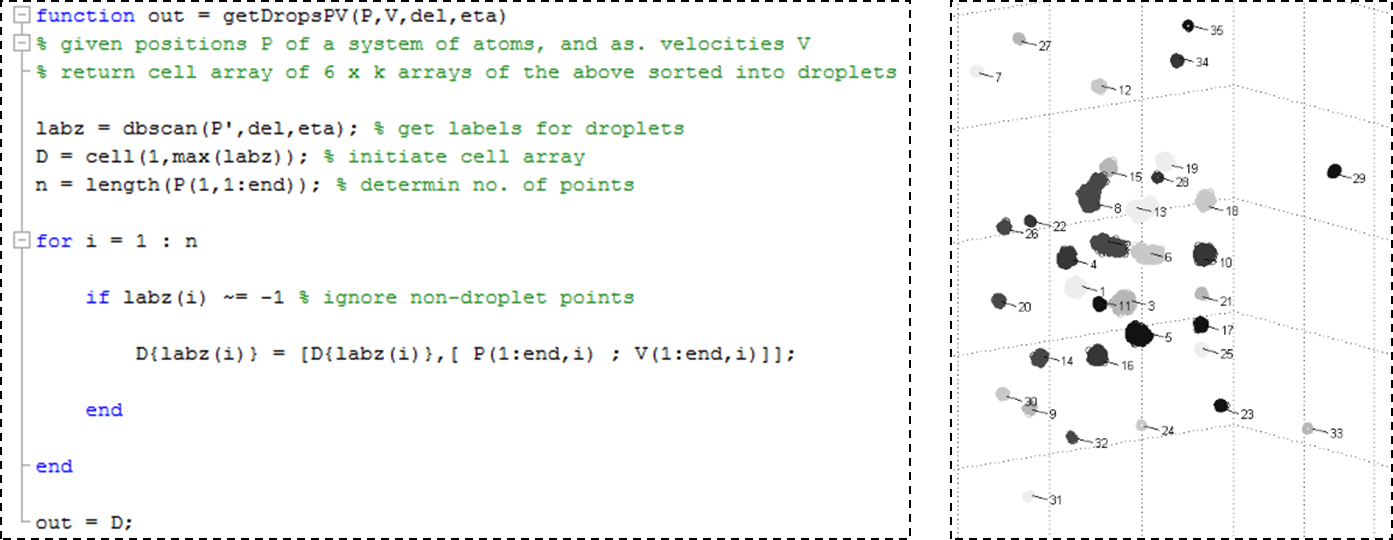}

$\;$

\textit{fig.24} - using DBSCAN to ID droplets (left), visual output (right) shows basic tagging \\

\end{center}

The variables \textit{del} and \textit{eta} above put into practice the $\delta$ and $\rho$ parameters respectively, from the second (more general) droplet definition given in chapter one. In the paper introducing DBSCAN \cite{bib:dbscan}, and in the algorithm's code, variables \textit{Eps} and \textit{MinPts} correspond to $\delta$ and $\rho$ respectively. The author apologizes for this inconsistency.\\

The purpose of identifying individual droplets visually was to give a means of isolating clusters, based on how well they conformed to being spherical in shape. This provided a basic means of testing the script \textit{sphDrop.m}, which implements the test of spherical measure presented at the end of the first chapter. See fig. 25 for examples of isolated droplets and their corresponding spherical measure ($\mu$) by this function.

\begin{center}
\includegraphics[scale=0.55]{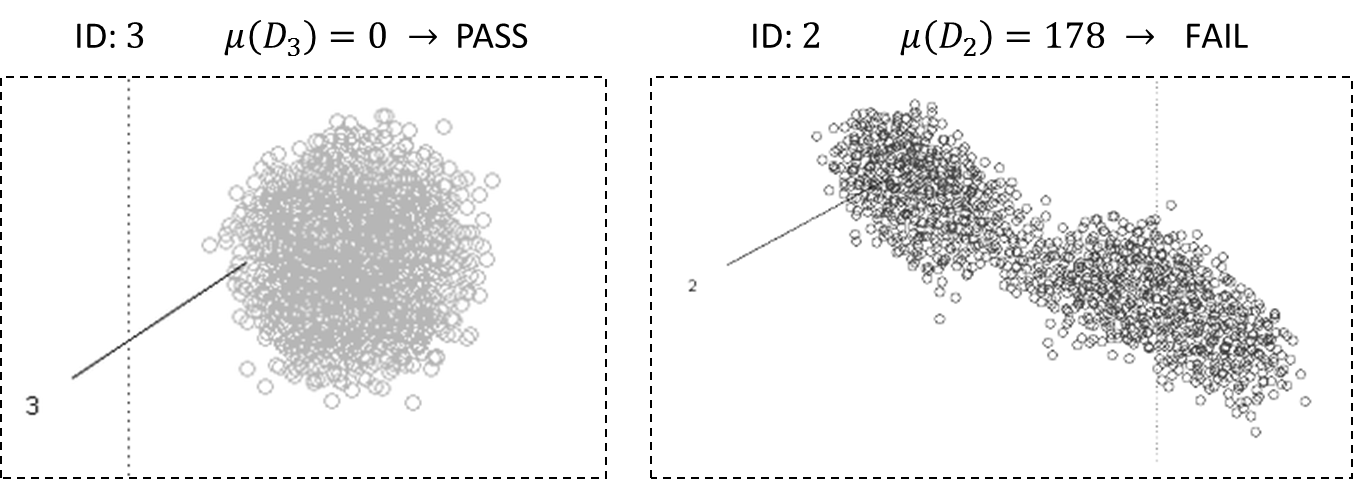}

$\;$

\textit{fig.25} - using tagged droplets for testing \textit{sphDrop.m}\\

$\;$

\end{center}

\subsubsection[OLS Algorithm]{\hyperlink{c38}{OLS Algorithm}}

In the analysis to come, it was required to perform basic fitting for a set of 2 dimensional data in order to test hypothesized correlation. This task is accomplished with relative ease through the use of MATLAB's built-in `Basic Fitting' toolbox; employing the (ordinary) linear least squares method to determine the unique straight line that minimizes the sum of vertical distances between data points and the line.\\

This approach isn't helpful, however, for automated correlation checking over many time-steps. To do this, the author used an existing script (\textit{leastsqrexp.m}, see fig. 26) implementing the OLS method for some given basis functions, written as part of the degree module `Scientific Computing' taken in 2013 and convened by the supervisor.

\begin{center}
\includegraphics[scale=0.55]{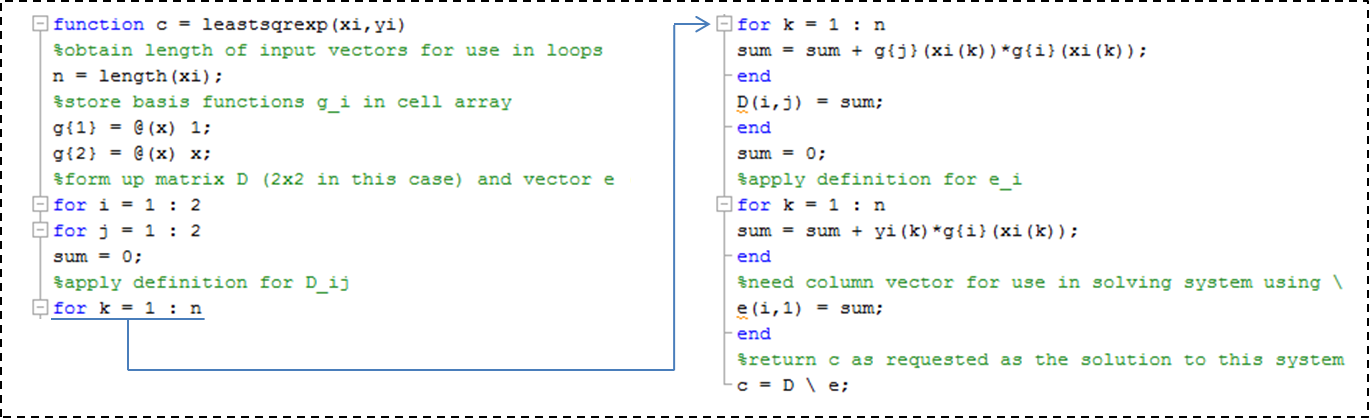}

\textit{fig.26} - \textit{leastsqrexp.m} code, finding coefficients for basis functions in OLS data fitting

\end{center}

Given a set of data points and basis functions $f_1(x),\cdots,f_n(x)$, this code returns coefficients $c_1, \cdots, c_n$ such that the function $F(x)=c_1f_1(x)+\cdots+c_nf_n(x)$ best fits the data in the sense of OLS. This is especially useful as it is the \textbf{gradient} of the fitting line that is of interest later. The recipe used for this code is taken from lecture notes \cite[pg.120]{bib:pav}.

\subsection[Model Testing and Finalization]{\hyperlink{c39}{Model Testing and Finalization}}

In this final section on implementation, checking of model validity and behaviour is demonstrated, and the penultimate steps leading to acquisition of main results are presented.


\subsubsection[Energy Exchange]{\hyperlink{c40}{Energy Exchange}}

A basic check for model accuracy is to observe that conservation of energy is obeyed (at least approximately). The function \textit{writeEnergies}, presented earlier, produces a file \textit{energies.txt} of total kinetic and potential energy within the system at periodic time-steps (governed by \textit{photoAvg} input in early testing). This can then be read by the script \textit{plotEnergies.m} to graph these values, in addition to their sum (total energy), see fig. 27.

\begin{center}
\includegraphics[scale=0.56]{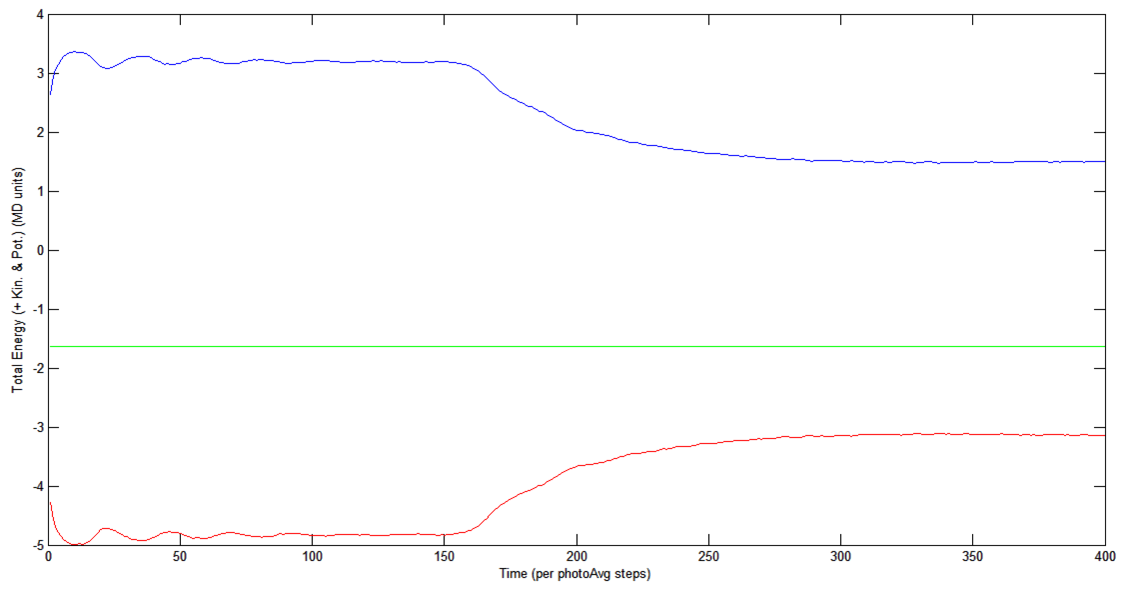}

\textit{fig.27} - observing kinetic (blue), potential (red) and total (green) energy

\end{center}

The \href{https://drive.google.com/file/d/0B1syDa_jHh0fNGVjakJ5VG5RSGc/view?usp=sharing}{plot above${}^{\star}$} demonstrates the general behaviour observed for all simulations checked (obeying PBC), in that there is some fluctuation initially, due to the initialization method, then a drop in kinetic energy (and necessary rise in potential) caused by the droplets colliding; the magnitude of the energy exchange being dependant on collision speed.

\subsubsection[Temperature Monitoring]{\hyperlink{c41}{Temperature Monitoring}}

In order to justify the state of the droplets pre-collision, to support the claim that they are liquid, it was necessary to monitor the temperature of a travelling droplet, using the definition provided in chapter 1. To do this (prior to being able to identify droplets), the heuristic function \textit{writeTemp} was included to the simulation code, writing temperature for the first half of the atoms in simulation (belonging to one of the two initial droplets); the results this produced are illustrated in fig. 28.

\begin{center}
\includegraphics[scale=0.54]{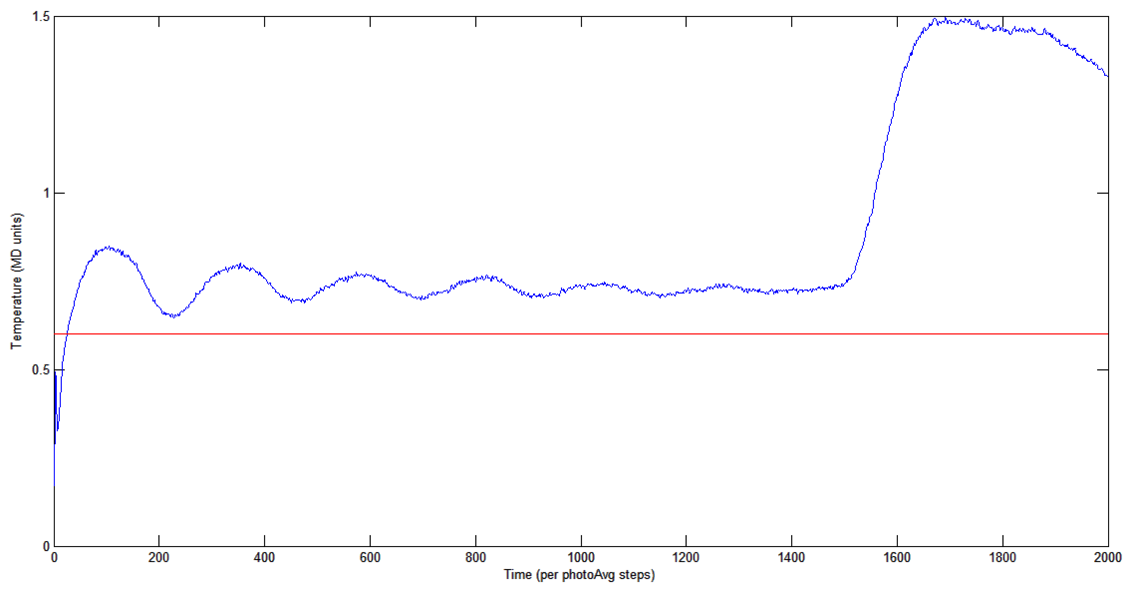}

\textit{fig.28} - monitoring temperature of a travelling droplet, comparing with melting point 0.6

\end{center}

The melting point value for this model was provided to the author as 0.6 (in MD units), and comparison with output temperature suggested that the droplets being simulated have temperature levels equilibrating to above this value prior to collision. Note again the fluctuation caused by the initial state, and that travelling droplets initially heat up. 

\subsubsection[Maximum Force Tests]{\hyperlink{c42}{Maximum Force Tests}}

It was realized early on in the project that a poor choice of time-step value ($\Delta t$) would completely invalidate the model, as the accuracy of solution provided by the numerical method can be disastrously poor if this parameter is too big. To investigate this issue and check the numerical method, the maximum force value (\textit{fcVal} in code) was measured across a number of time-step values, see fig. 29 for results.

\begin{center}
\includegraphics[scale=0.58]{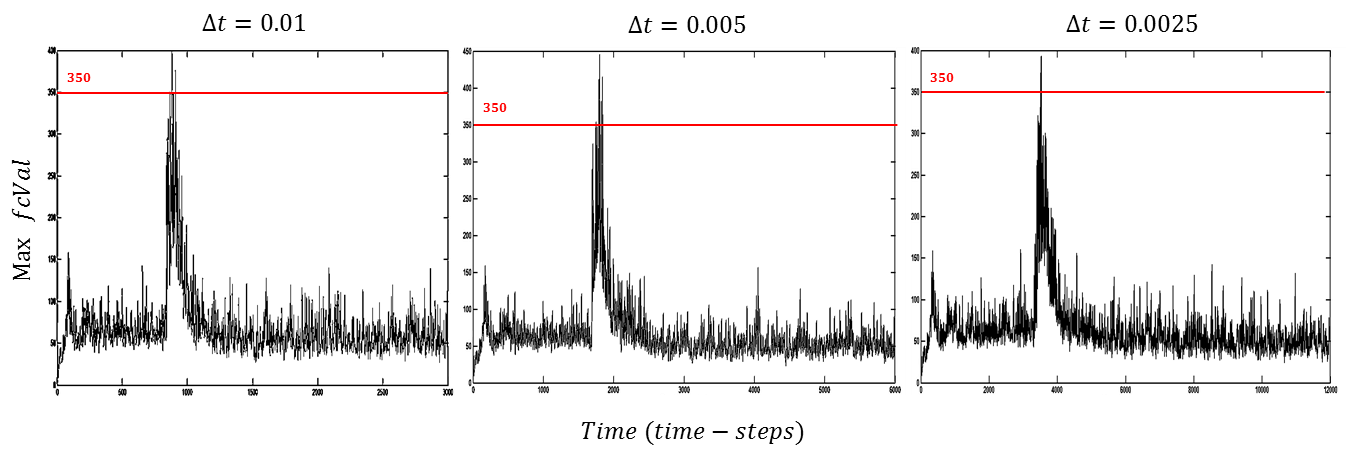}

\textit{fig.29} - testing numerical method: observing consistency in max. force, varying $\Delta t$

\end{center}

\subsubsection[Collision Region Dimensions]{\hyperlink{c43}{Collision Region Dimensions}}

The collision region used throughout most of the project was cubic, a natural choice since initially there was no incentive to make it irregular. However, constraints on region size (caused by cell subdivision cost), combined with a desire to make the region volume as large as possible, motivated a reconsideration of this decision.\\

A feature observed in watching head on collisions of droplets is that the density of atoms post-collision appeared highest around the plane orthogonal to the collision trajectories. In this project, droplets approach each other along the $x$-axis and collide about the origin, and so the bulk of post-collision atoms favour the vicinity of the $z,y$-plane, see fig. 30 for illustration ($z,y$-plane is shaded blue).
$\;$\\
\begin{center}
\includegraphics[scale=0.54]{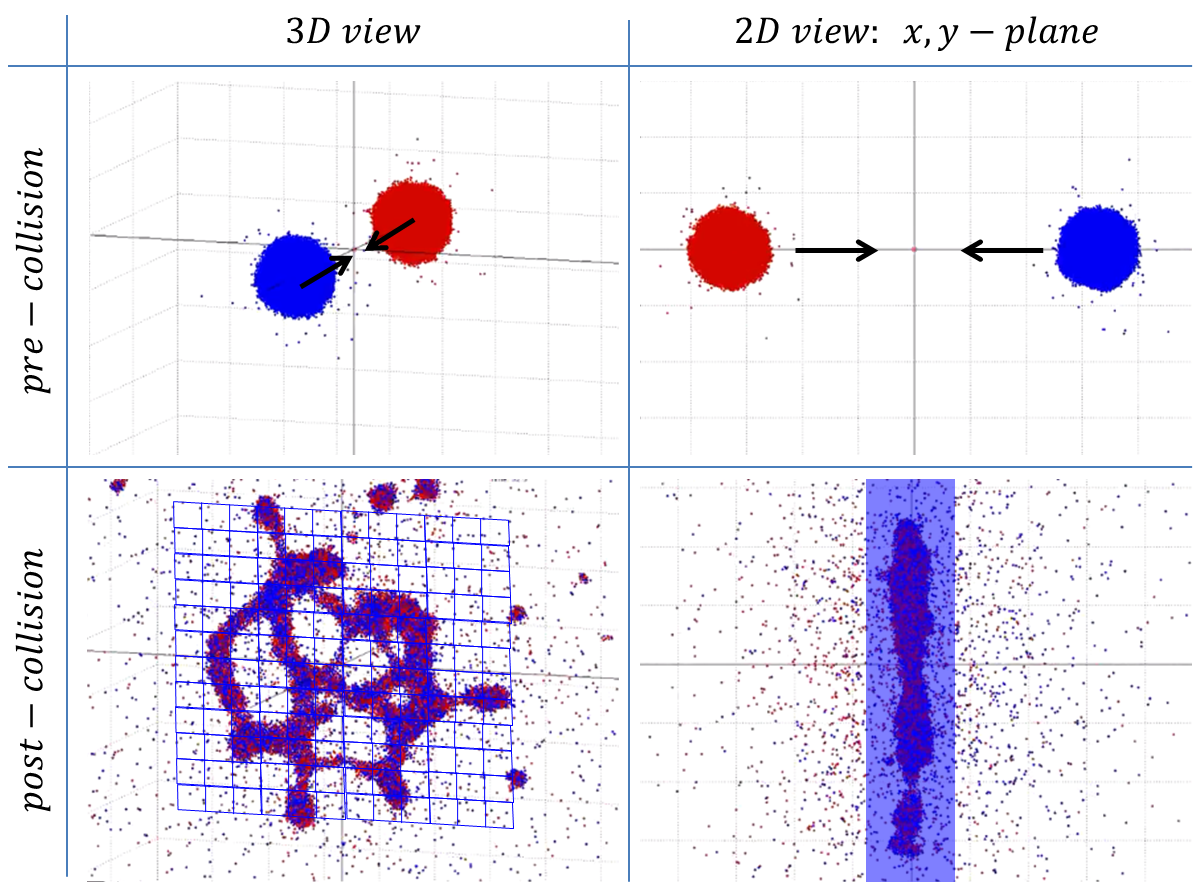}

\textit{fig.30} - \href{https://drive.google.com/file/d/0B1syDa_jHh0fTVVWYl9sZzkybUU/view?usp=sharing}{illustrating collision debris favouring plane orthogonal to approach axis${}^{\star}$}

$\;$\\

\end{center}

Thus it was decided to implement a non-cubic cuboid region with side lengths $R_x,R_y,R_z$ in the $x,y,z$ directions respectively, obeying $R_y=R_z$ and $R_x < R_y,R_z$, to take advantage of the observed collision behaviour and optimize the volume being simulated. The important outcome of this realization is that, for fixed region volume, post-collision droplets can be tracked for longer before they are lost to the boundaries.

\subsubsection[Main Simulations]{\hyperlink{c44}{Main Simulations}}

It was necessary at some point in the simulation development, a seemingly endless maze of opportunity for improvement, to settle on a particular set of parameters and partition of collision speeds on which to perform the main simulations for analysis.\\

The author chose to run simulations over \textbf{10} collision speeds, $s(D)= 1 , 1.25 , \cdots , 3 , 3.25$ (MD units). To provide some justification for this choice, it was suggested that there would be two extremes of collision behaviour, a single post-collision droplet (low speed) and many fast-forming post-collision droplets (high speed), and from experiment this partition was deemed adequate in covering this range (see fig. 31).

\begin{center}
\includegraphics[scale=0.5]{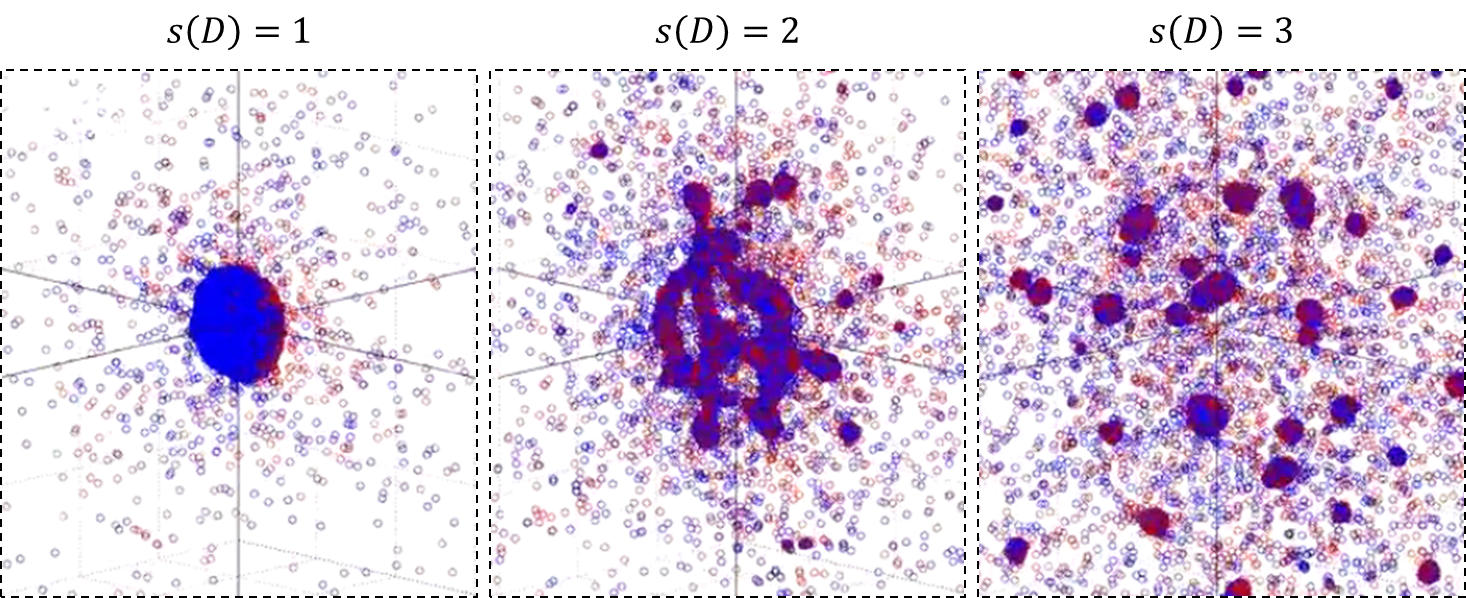}

\textit{fig.31} - demo. qualitative difference between \href{https://drive.google.com/file/d/0B1syDa_jHh0fdGxjVG5kSkgyRkU/view?usp=sharing}{iconic speeds${}^{\star}$} (slow, medium, fast)

\end{center}

Given a desired collision speed $s(D)$, the $\alpha$ and $\beta$ parameters introduced in describing the droplet initialization algorithm (chapter 1), governing randomness of initial atom speed, are set to $(\alpha,\beta)=(s(D)-0.05,s(D)+0.05)$. All other parameters are the same across all speeds, and are provided in the following table (fig. 32). 

\begin{center}
\includegraphics[scale=0.6]{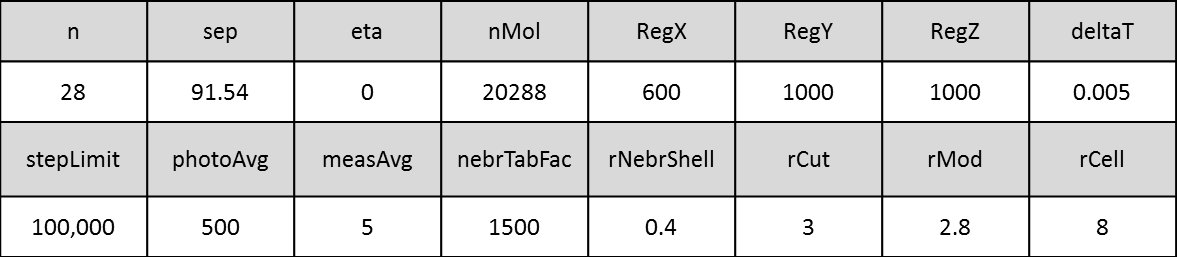}

\textit{fig.32} - table of parameters used in the main simulations (aids reproducibility)

\end{center}

N.B. parameters \textit{rNebrShell} and \textit{nebrTabFac} govern properties of the neighbour list method employed by Rapaport, covered from page 54 of his book \cite{bib:rapa}.

\subsubsection[Analysis Scripts]{\hyperlink{c45}{Analysis Scripts}}

Once the main simulations had been run, it became clear (towards the end of the project) that there were three major areas of interest for analysis, these being \textbf{size}, \textbf{speed} and \textbf{temperature}. A number of preliminary observations performed statically (individual time-steps on certain speeds) persuaded the author to perform speed analysis on only the last 4 speeds, utilizing the spherical measure, and perform analysis of the other two properties on all 10 speeds without spherical measure.\\

Thus two MATLAB scripts, \textit{sizespeedcorr.m} and \textit{sizetempcorr.m} (see fig. 33), were written for submission to ALICE, carrying out the expensive task of droplet analysis, over many time-steps and across multiple speeds. The expensiveness is primarily due to the cost of identifying droplets from the given set of over 20,000 points, in addition to some difficulty getting MATLAB to execute efficiently on ALICE.

\begin{center}
\includegraphics[scale=0.7]{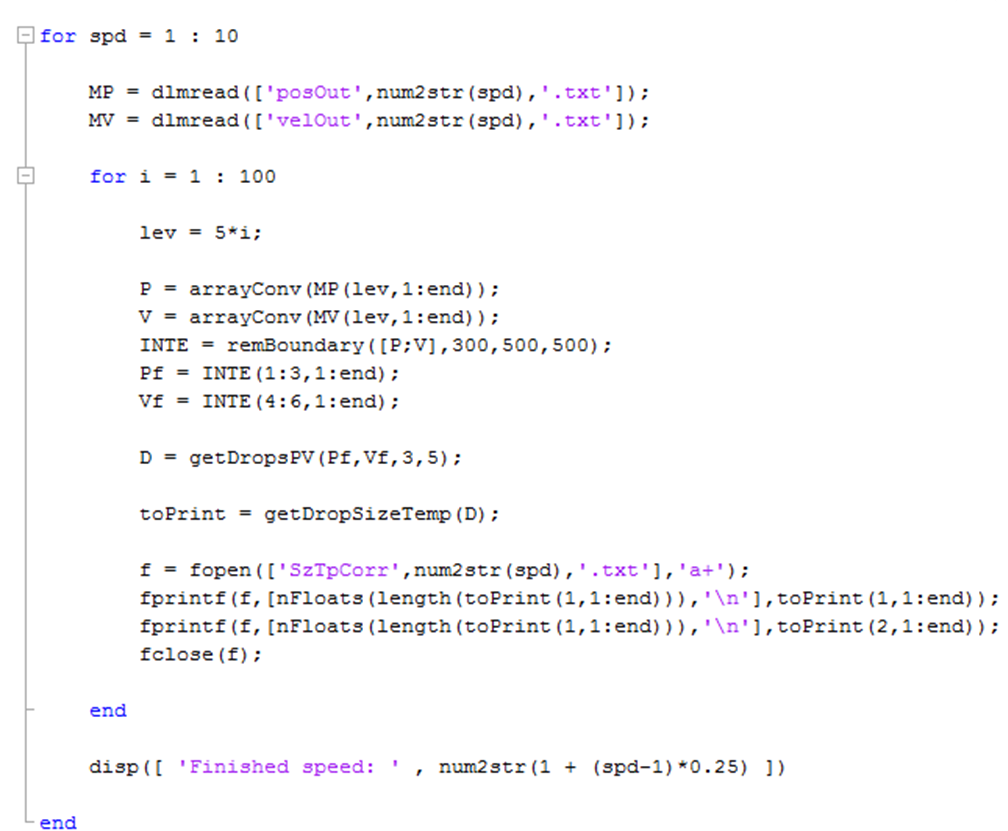}

\textit{fig.33} - script processing output positions and velocities over all $s(D) \in \{1,\cdots,3.25\}$

\end{center}

The result of running these scripts is a set of .txt files containing the post-processed data required for large scale droplet analysis (i.e. observing properties across collision speeds and many time-steps), the results of which are discussed in chapter 3. The longer \textit{sizespeedcorr.m} script is provided as \hyperlink{A4}{item 4${}^{\star}$}\hypertarget{A4r}{} in the appendix, and implements a basic (and successful) collision detection algorithm, so that speed is primarily recorded post-collision.


\section[Results]{\hyperlink{c46}{Results}}

In this final chapter, some basic analysis of data obtained, from the 10 simulations previously described, is presented. The nature of this analysis is largely qualitative and observational, due partly to the amount of work taken to get here, i.e. effort expended on model development left little time for in-depth analysis.

\subsection[Droplet Size-Speed Correlation]{\hyperlink{c47}{Droplet Size-Speed Correlation}}

Though the initial interest in this project was to study the dependence of post-collision droplet size distribution on the collision speed, the focus mid-project was shifted to the droplet size-speed relation; as the supervisor recognized a correlation from visual output, that smaller droplets were travelling faster post-collision, an interest further bolstered by a suggested dependence on droplet radius (Prof. Brilliantov).

\subsubsection[Hypothesized Relationship]{\hyperlink{c48}{Hypothesized Relationship}}

The hypothesis to be tested is: for a droplet formed post-collision, its speed is inversely proportional to its radius. So given a droplet of $N$ atoms (measuring its volume), with centre of mass velocity $v_c$, giving droplet speed $s = |v_c|$, the following is assumed:

$$s \propto \frac{1}{\sqrt[\leftroot{0}\uproot{2}3]{N}} \;\;\;\; \Leftrightarrow \;\;\;\; s = \kappa N^{(-1/3)} \;\; , \;\; \kappa \in \mathbb{R}_+ \;\;\;\; \Leftrightarrow \;\;\;\; \log(s) = -\frac{1}{3}\log(N)+\log(\kappa)$$
$\;$

Thus this assumption predicts that plotting $\log(s)$ against $\log(N)$ for a number of post-collision droplets should give a straight line with gradient adhering to $(-1/3)$. Of course the reality, given a large number of approximations, is not so neat; however linear fitting can still be conducted (previously eluded to) to check this relation.

\subsubsection[Encouraging Results]{\hyperlink{c49}{Encouraging Results}}

To begin with, a number of single time-steps from various high speed collisions were examined, of which one is presented here (fig.s 34, 35). In all cases there was a clear negative correlation between droplet size and speed, as predicted.

\begin{center}
\includegraphics[scale=0.52]{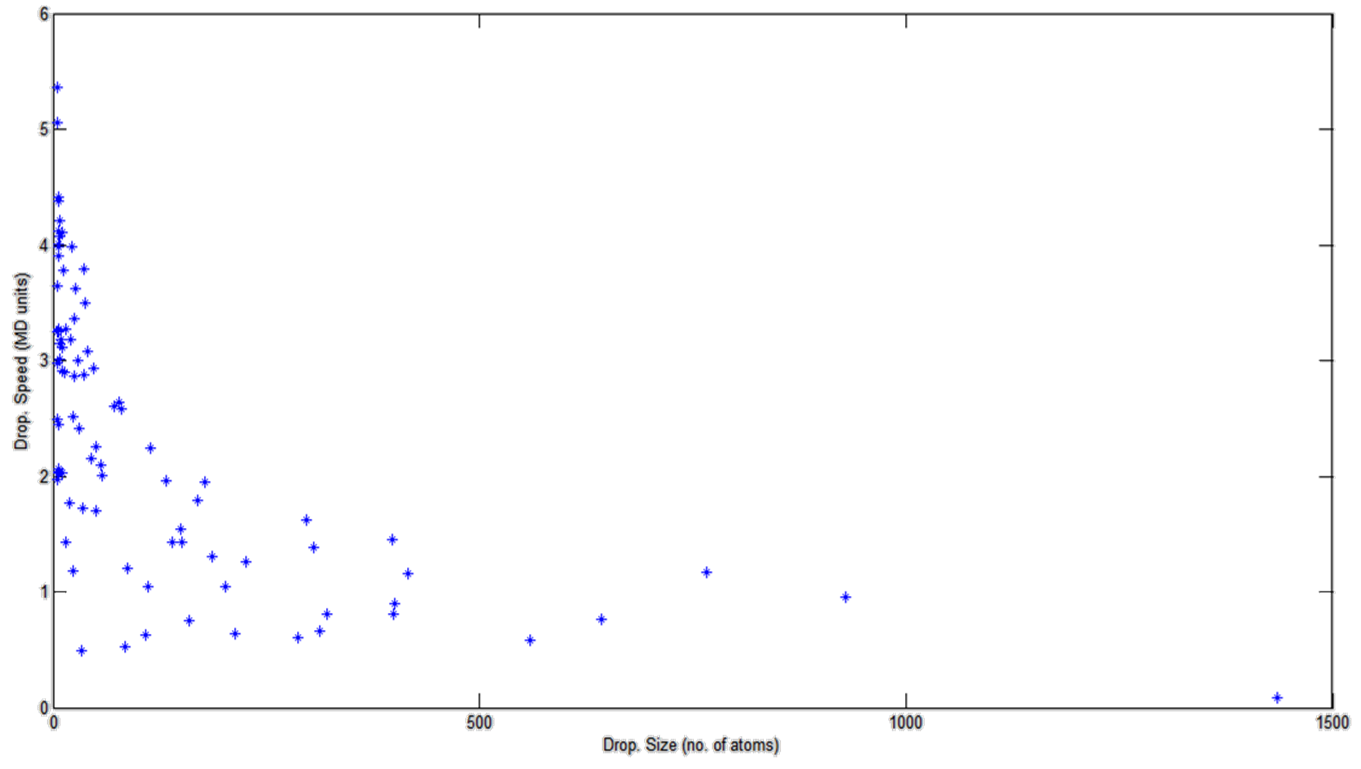}

\textit{fig.34} - droplet size-speed correlation, $s(D)=3$, time-step 70,000

\end{center}

Taking logs and applying linear fitting (using OLS) to these results often produced a gradient of close to $-1/3$, as the example in fig. 35 demonstrates (gradient read from given equation is $-0.33206$). In general this appears to be the case for ideal time-frames (close to impact with many spherical droplets).

\begin{center}
\includegraphics[scale=0.48]{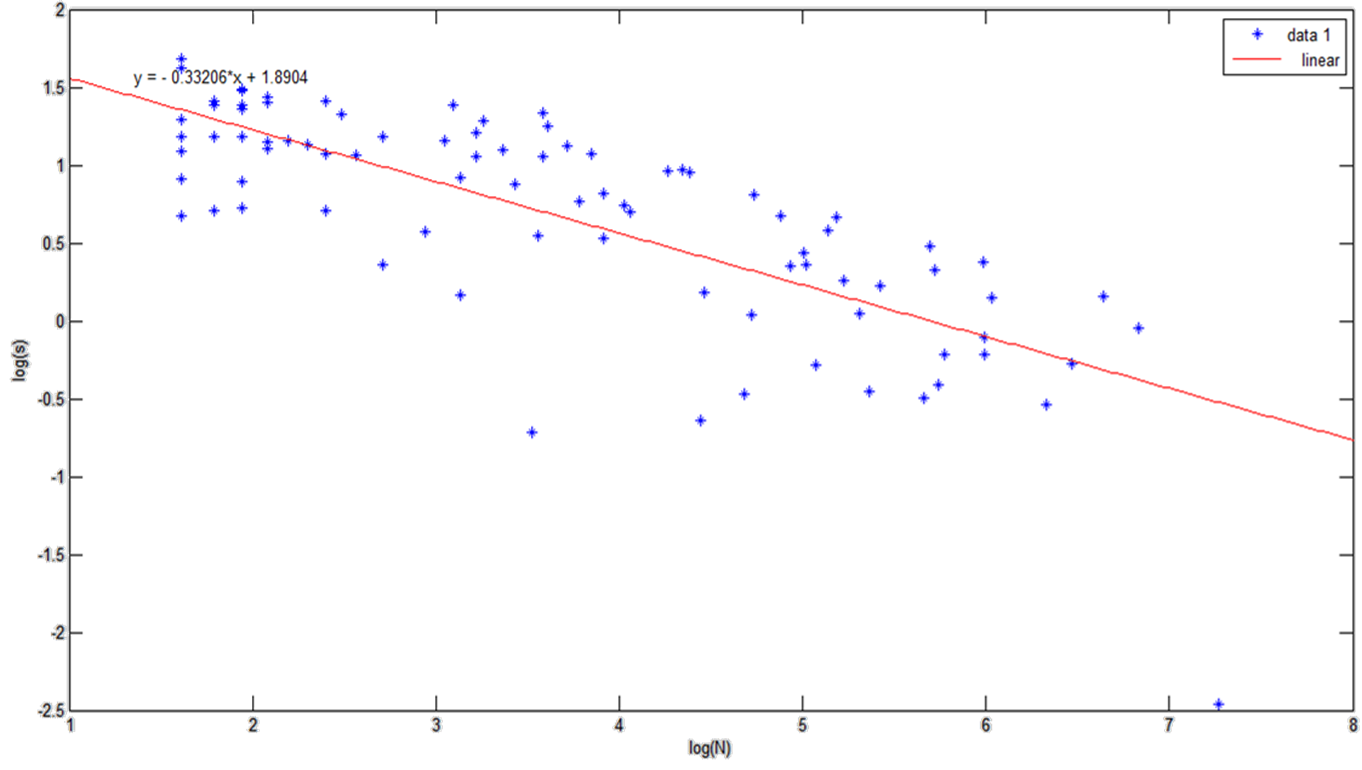}

\textit{fig.35} - loglog plot of data from fig. 34, testing hypothesised relationship

\end{center}

To provide a general view of these results, the aforementioned script \textit{sizespeedcorr.m} was executed examining the highest 4 collision speeds, of which speed 2.5 is demonstrated in fig. 36. Across all 4 speeds it was observed that the relevant OLS coefficient tends to around $(-1/3)$ soon after impact, then increases some time after, with magnitude of diversion depending on collision speed (higher speed diverges faster and to a greater extent).

\begin{center}
\includegraphics[scale=0.5]{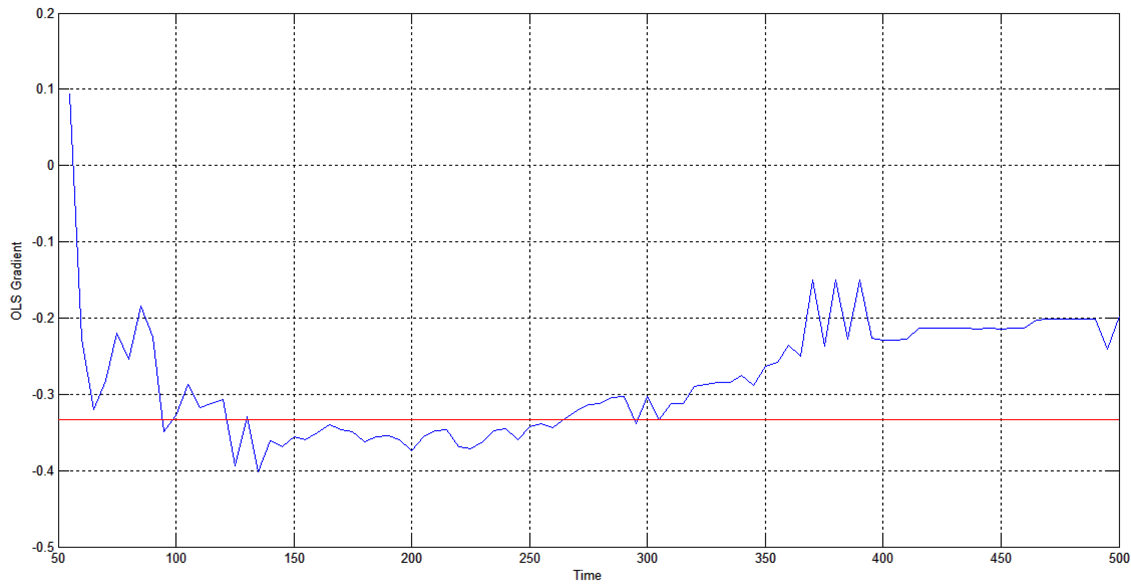}

\textit{fig.36} - automated droplet sz-spd corr. results (\textit{sizespeedcorr.m}) for $s(D)=2.5$

\end{center}

It was realized that the implemented translational boundary conditions, combined with a region too small to contain all droplets for the duration of simulation, was having an undesired effect on the results. In addition to loosing droplets at the boundaries, the method of extraction (atom-by-atom) caused fragmentation that could add further distortion.\\

This flaw does not, however, invalidate observations made immediately post-impact (takes time for droplets to reach the boundary and incur error). An attempt to study this issue will be presented later in looking at average droplet size over time, where system vapour (atoms failing the \textit{minPts} parameter in DBSCAN) and vacant atoms are counted.

\subsection[Droplet Size-Temperature-Time Correlation]{\hyperlink{c50}{Droplet Size-Temperature-Time Correlation}}

The supervisor expressed an interest in the distribution of temperature in the droplets post-collision, and so motivated investigation of droplet size-temperature correlation. As before, this began with examining a particular point in time for one collision speed, then a more general picture is provided via 3D plots, comparing size, temperature and time over multiple collision speeds. In addition, average droplet size for varying collision speeds is presented with vapour and vacant atom counting.

\subsubsection[Size-Temperature Observations]{\hyperlink{c51}{Size-Temperature Observations}}

The graph provided in fig. 37 demonstrates a positive correlation between droplet size and temperature post-collision, that is, larger droplets appear to retain more temperature than smaller ones. The red line plotted marks the melting point of 0.6 for this model, and the magenta curve is an attempt by the author to perform basic fitting using the OLS script \textit{leastsqrexp.m} with basis functions $1$ and $log(x)$.

\begin{center}
\includegraphics[scale=0.44]{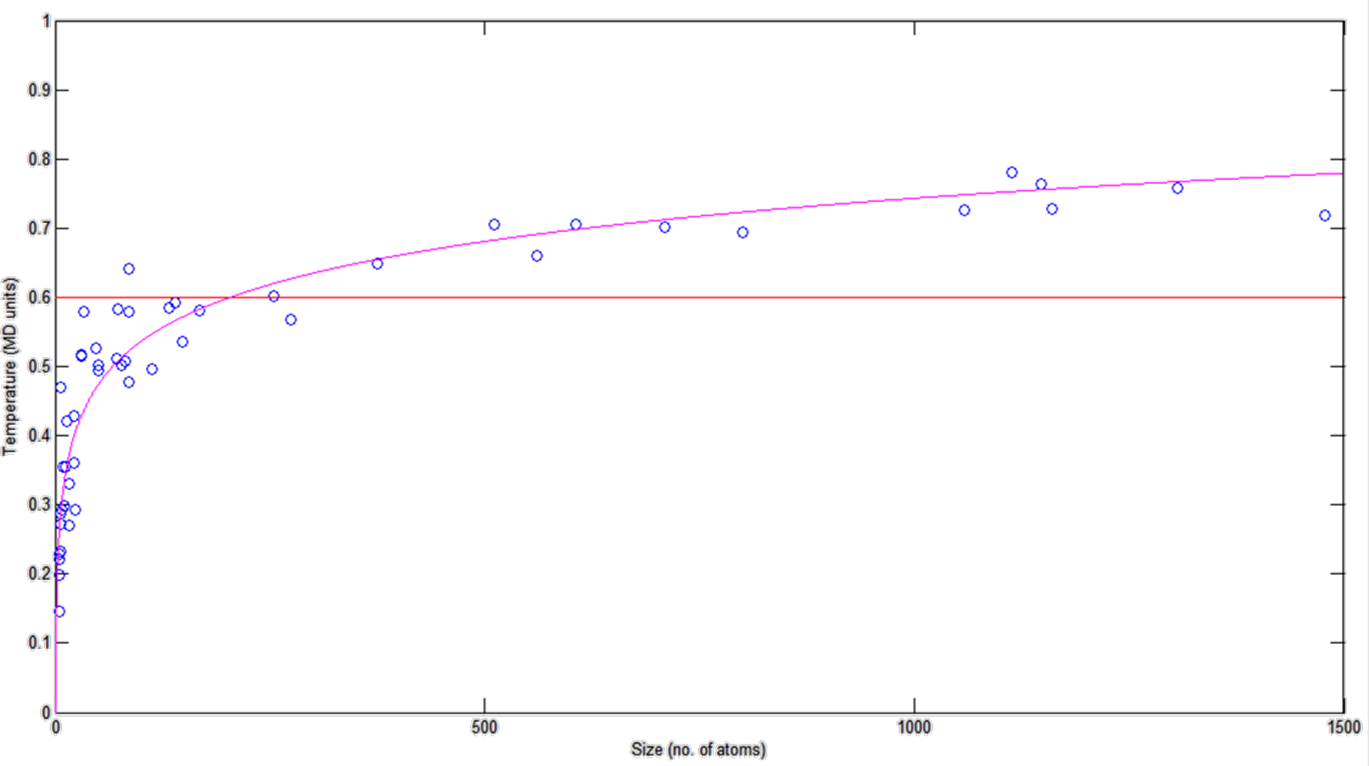}

\textit{fig.37} - size-temp. corr. with `log' fitting, time-step 70,000, $s(D)=2.5$

\end{center}

This positive correlation permeates all post-collision time-steps for all collision speeds, see fig. 38 for demonstration with collision speed 2.5. It was also observed that higher collision speed leads to a higher droplet temperature, particularly in the larger droplets, immediately post-impact, illustrated in fig. 39.  \\
$\;$

\begin{center}
\includegraphics[scale=0.4]{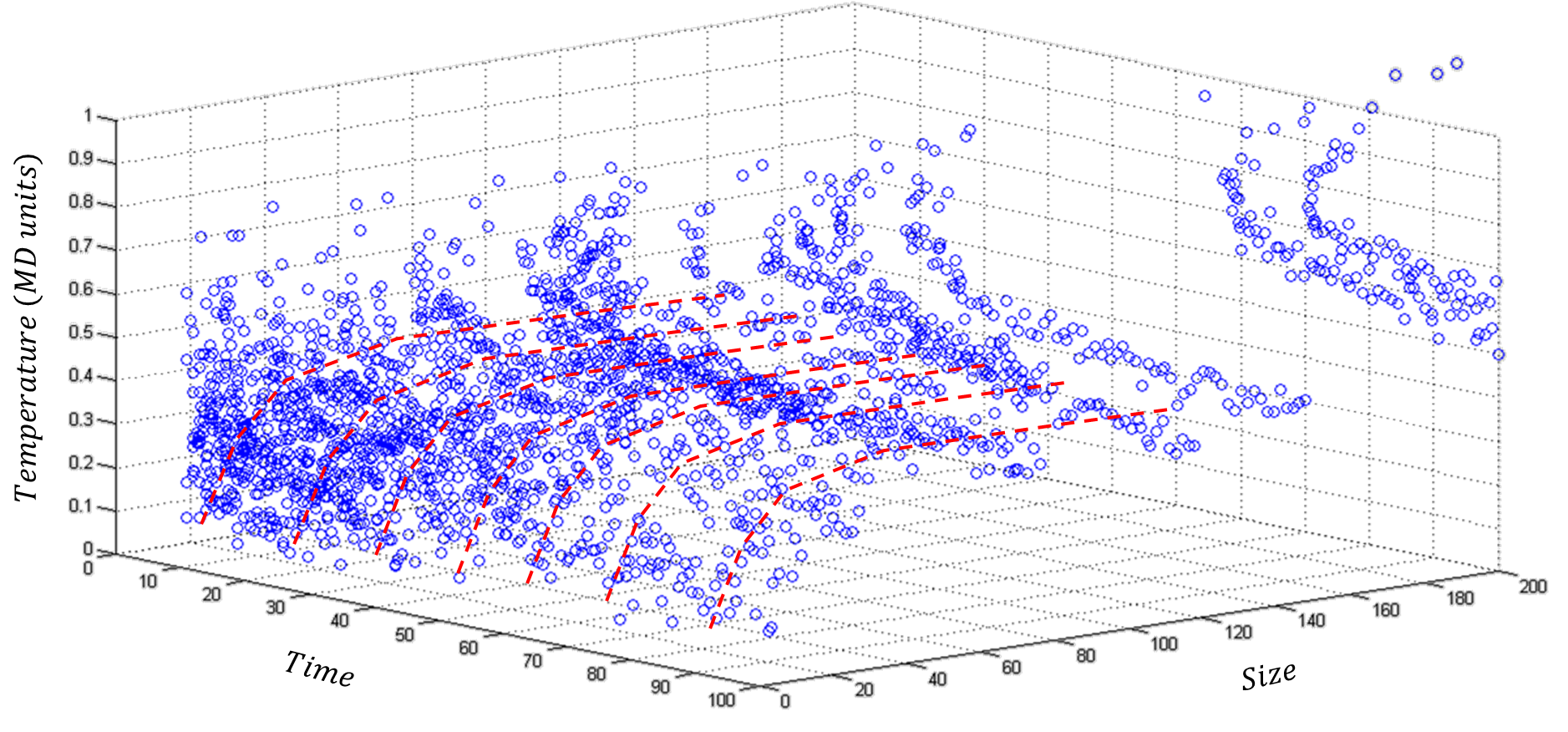}

\textit{fig.38} - observing size-temp. dependence over all time-steps for $s(D)=2.5$

\end{center}

\begin{center}
\includegraphics[scale=0.38]{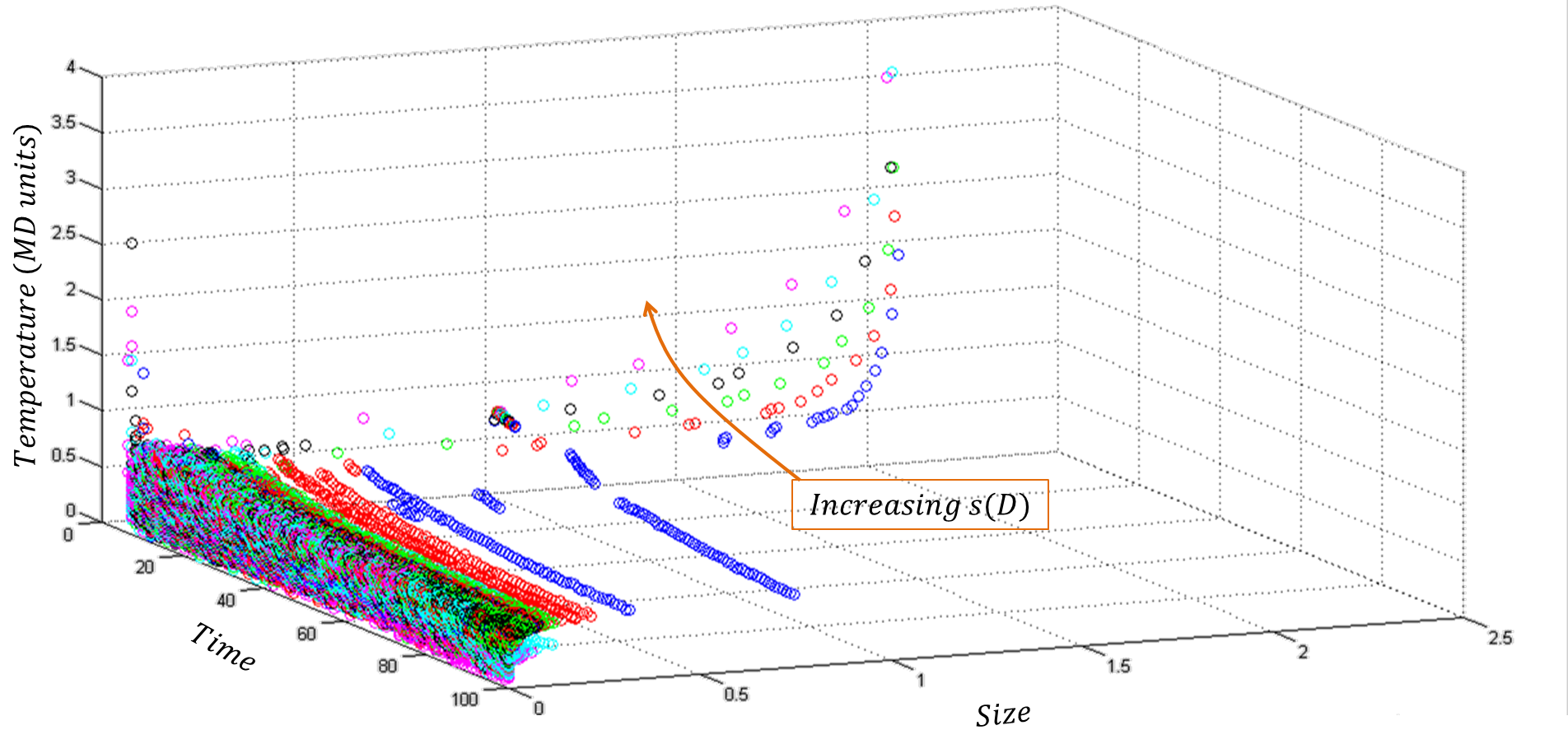}

\textit{fig.39} - observing difference in immediate post-impact drop. temp., $s(D)=2,\cdots,3.25$

\end{center}

\subsubsection[Size-Time Observations]{\hyperlink{c52}{Size-Time Observations}}

The original goal of this project was to study the dependence of post-collision average droplet size on collision speed. In practise this proved difficult to examine due to interference from boundary conditions and limited region size. However, some interesting qualitative behaviour can still be observed, and investigation of this issue is presented to indicate the extent of negative effects on the results.

\begin{center}
\includegraphics[scale=0.42]{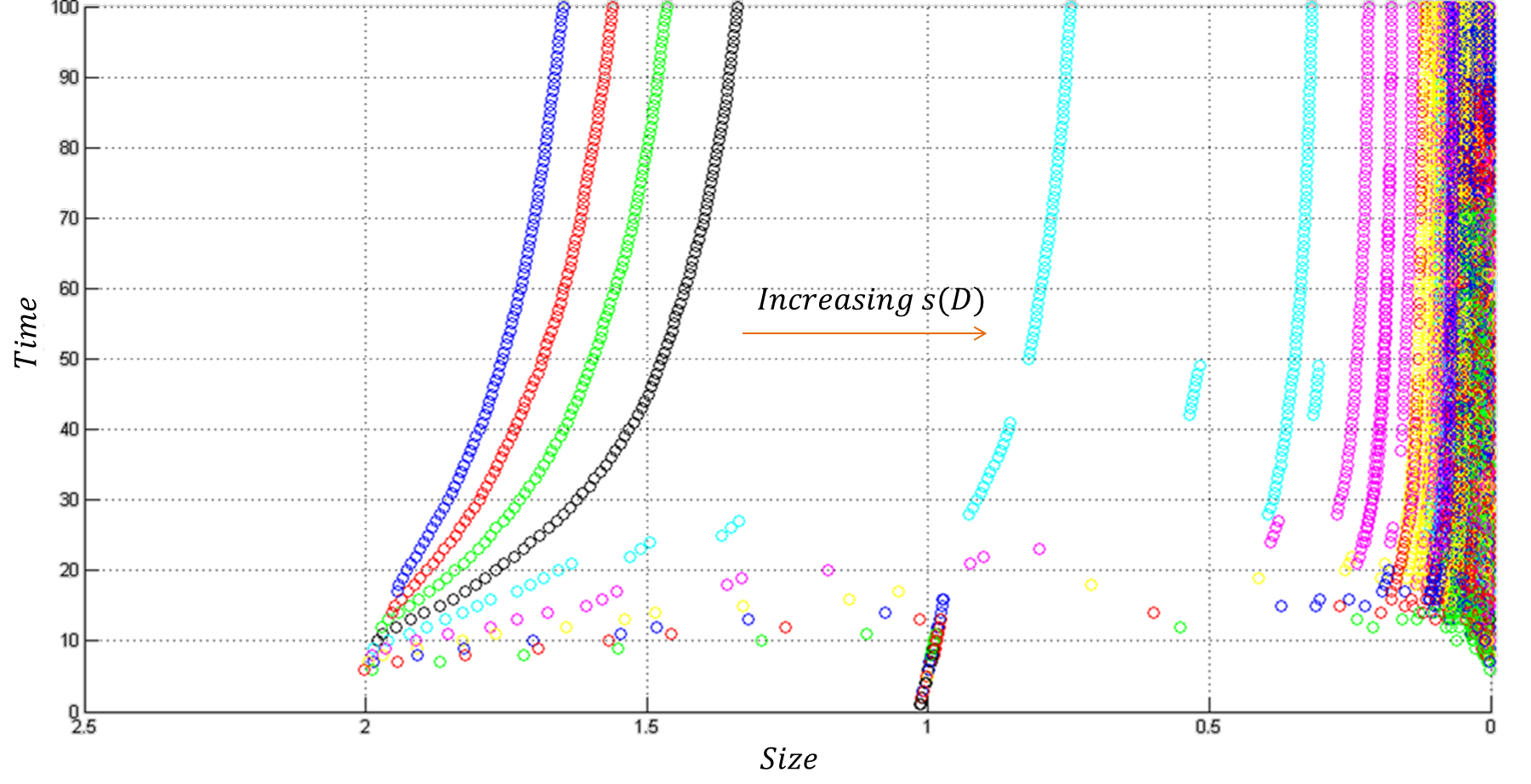}

\textit{fig.40} - qualitative size analysis in time for all 10 speeds

\end{center}

The image above (fig. 40) visually demonstrates some interesting features. There is a clear qualitative jump between $s(D) \in [1,1.75]$ and higher speeds, i.e. between one droplet post-collision and many. The speed $s(D)=2$ appears to be close to the transitional point. Evaporation can be appreciated for $s(D) \in [1,1.75]$, indicating higher $s(D)$ gives a larger gradient immediately post-impact, but in time the gradients become approximately equal.

Change in average droplet size over time, whilst recording vacating atoms and vapour for error indication, is demonstrated in fig. 41. The graphs illustrate, to some extent, the reliability of results for varying $s(D)$, measured by the vacated atom count. In particular, these show that, even for the highest speed simulated $s(D)=3.25$, the immediate post-impact time frames are relatively unaffected. \\

\begin{center}
\includegraphics[scale=0.46]{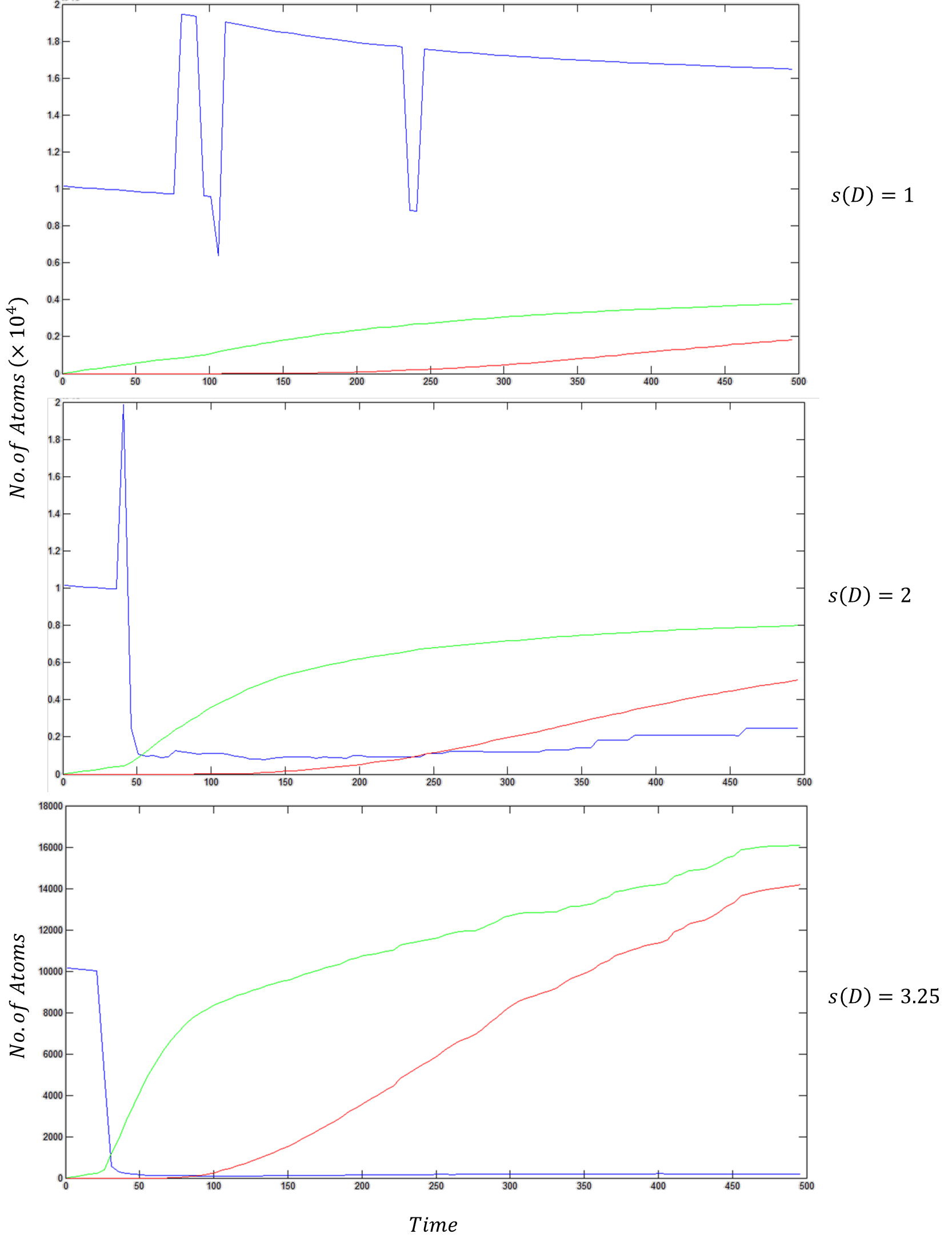}

\textit{fig.41} - plotting avg. drop. size (blue), vapour in system (green), and vacated atoms (red)

\end{center}

\subsection[Closing Remarks]{\hyperlink{c53}{Closing Remarks}}

The aim of this project was to model and study head-on liquid He nanodroplet collision using MD simulation techniques, which the author feels has \href{https://drive.google.com/file/d/0B1syDa_jHh0fMUZsd3FoTENpSTg/view?usp=sharing}{clearly been achieved${}^{\star}$} (admittedly resigning the He aspect to L-J generalization). Beyond this, a great deal of unanswered questions have arisen that encourage investigation.\\

A prominent issue, given flaws presented previously in the model design, is how best to deal with atoms, and more importantly droplets, exiting the simulation region. The boundary conditions discussed are inappropriate for reasons given, and making the region size sufficiently large appears terminally expensive to compute.\\

One evident possibility is to perform droplet analysis within the simulation (i.e. as the simulation is running), with an aim to identify when a droplet is about to hit a boundary, then remove that entire droplet whilst recording its properties (positions and velocities of constituent atoms). If the cost of identifying droplets in simulation isn't too large (as discussed, it isn't particularly cheap), such an implementation in a relatively small simulation region would be ideal for analysing formed-droplet properties post-collision.\\

Another avenue of investigation suggested to the author, though too late to work on, was the idea of studying droplet homology, in particular the cluster genus (number of holes roughly speaking, see \hyperlink{A2}{item 2${}^{\star}$}\hypertarget{A2r}{} in appendix) in collision deformation. It would be interesting to implement an algorithm capable of accurately measuring this property, to test for correlation with other system properties (in particular, no. of droplets formed).

\begin{center}
\includegraphics[scale=0.61]{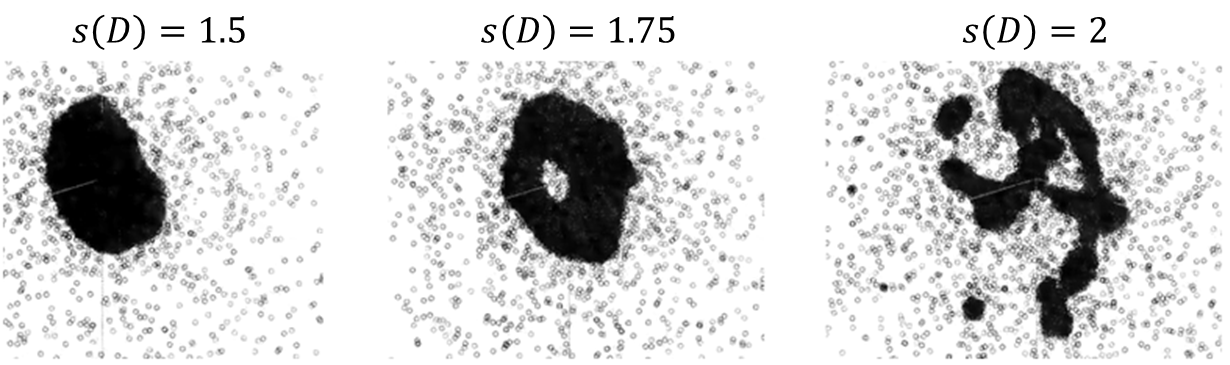}

\textit{fig.42} - observing qualitative behaviour across transition (one drop. to many) boundary

\end{center}

Finally, and relating still to droplet homology, it would be interesting to study the dividing line between `one' and `many' post-collision droplets. In particular, demonstrated in fig. 42, there appears a toroidal structure in $s(D)$ approaching this transition point. There is an obvious motivation to study the interval \href{https://drive.google.com/file/d/0B1syDa_jHh0fTldKdGpwM2NpVzA/view?usp=sharing}{$s(D) \in [1.75,2]$ ${}^{\star}$}.

\subsection*{Acknowledgements}

The author would like to acknowledge the following academics for their support of this project and  significant contribution to the work:

\begin{itemize}
	
	\item \href{http://www2.le.ac.uk/departments/mathematics/extranet/staff-material/staff-profiles/rld8}{Prof. Ruslan Davidchack${}^{\star}$} \textit{(Department of Mathematics, University of Leicester)}

	As project supervisor, provided guidance and suggestions throughout the project, and is responsible 		for its inception. Amongst many things, provided instruction for use of a switching function (modifying L-J potential), suggested cubic lattice method for droplet initialization, provided access to ALICE and instruction on its use.\\
	
	\item \href{http://www2.le.ac.uk/departments/mathematics/extranet/staff-material/staff-profiles/nb144}{Prof. Nikolai Brilliantov${}^{\star}$} \textit{(Department of Mathematics, University of Leicester)}
	
	Motivated the study of droplet size-speed correlation post-collision with suggested inverse relationship to be tested, inspiring work in this direction. Additionally provided relevant material and further suggestions for the author to review.\\
	
	\item \href{http://www2.le.ac.uk/departments/mathematics/extranet/staff-material/staff-profiles/ag153}{Prof. Alexander Gorban${}^{\star}$} \textit{(Department of Mathematics, University of Leicester)}

	Motivated the study of temperature and maximum force in the model, to better determine nature of droplets being simulated and accuracy of the numerical method.\\
	
	\item \href{http://www2.le.ac.uk/departments/chemistry/people/academic-staff/andrew_m_ellis}{Prof. Andrew Ellis${}^{\star}$} \textit{(Department of Chemistry, University of Leicester)}
	
	Provided advice and insight regarding the real nature of liquid / superfluid Helium. Demonstrated experimentation techniques involving liquid He nanodroplets.\\
	
	\item \href{http://www.ph.biu.ac.il/~rapaport/}{Prof. Dennis Rapaport${}^{\star}$} \textit{(Department of Physics, Bar-Ilan University)}
	
	Though not personally involved in the project, it is \href{http://www.ph.biu.ac.il/~rapaport/mdbook/index.html}{his book${}^{\star}$} \cite{bib:rapa} that formed the basis of the project work; and code made available by him that formed the basis of the simulation code (source scripts altered to meet project requirements).

\end{itemize}

\newpage

\begin{center}
\begin{huge}
\textbf{Appendix}
\end{huge}
\end{center}

$\;$

\begin{center}
\hypertarget{A1}{}
\textbf{Item 1} - \hyperlink{A1r}{phase transition diagram for stable Helium isotope ${}^4$He ${}^{\star}$}\\
Source: \url{http://ltl.tkk.fi/research/theory/He4PD.gif}\\

$\;$\\

\includegraphics[scale=0.8]{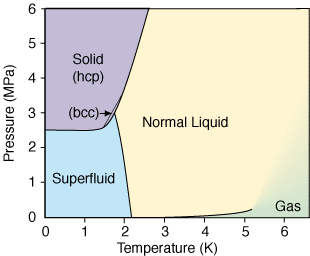}

\end{center}

$$\;$$

\begin{center}
\hypertarget{A2}{}
\textbf{Item 2} - \hyperlink{A2r}{image of collision genus in post-impact deformation${}^{\star}$}\\

$\;$\\

\includegraphics[scale=1]{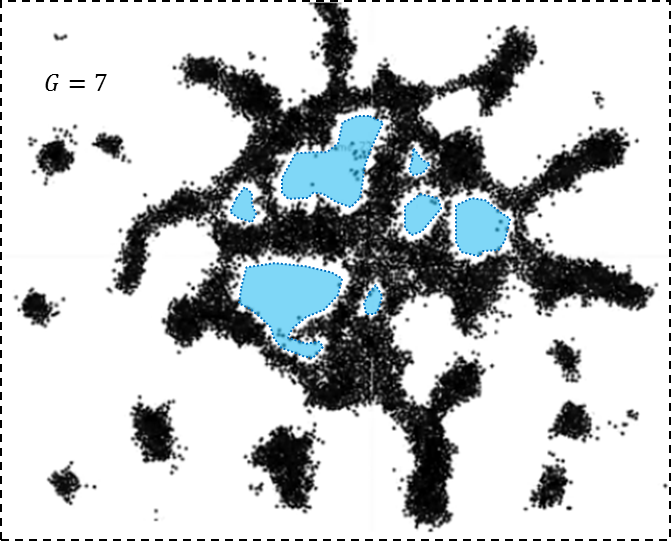}

\end{center}

\newpage

\begin{center}
\hypertarget{A3}{}
\textbf{Item 3} - \hyperlink{A3r}{algorithm for droplet initialization, generating initial state${}^{\star}$}\\

$\;$\\

\includegraphics[scale=0.8]{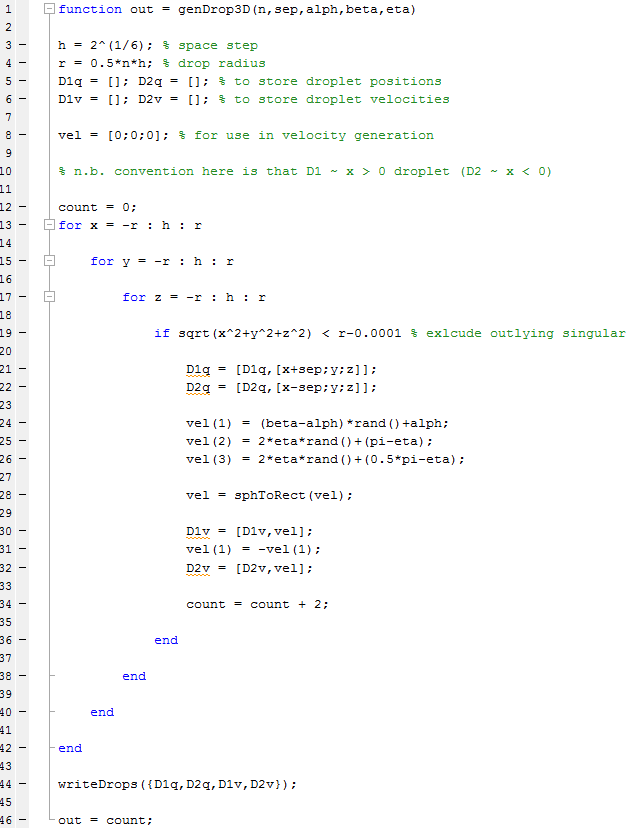}

\end{center}

 \newpage

\begin{center}
\hypertarget{A4}{}
\textbf{Item 4} - \hyperlink{A4r}{algorithm for size-speed analysis, including primitive collision detection${}^{\star}$}\\

$\;$\\

\includegraphics[scale=0.75]{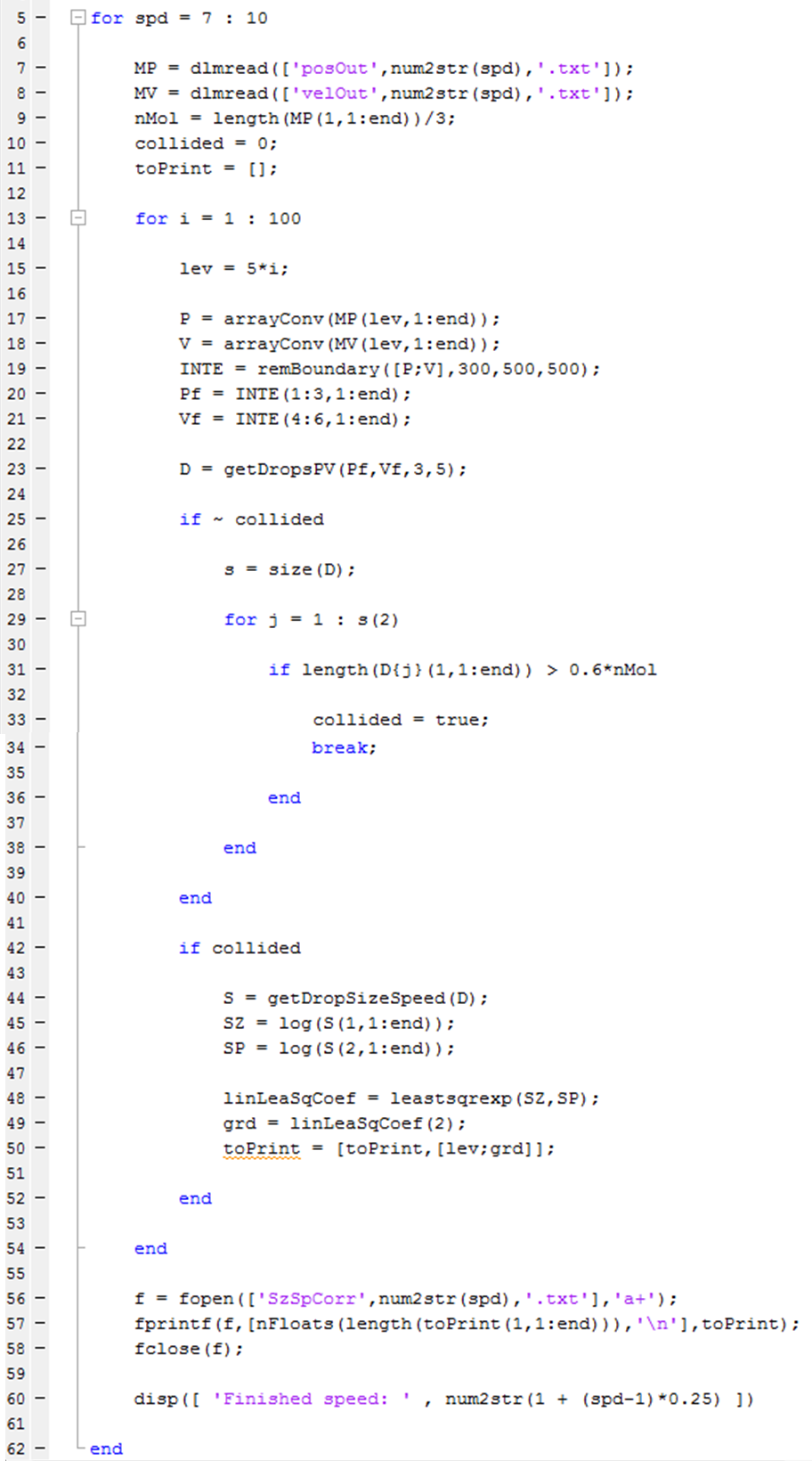}

\end{center}

\end{document}